\documentclass[numberedappendix]{emulateapj}

\begin{document}

\bibliographystyle{apj}



\slugcomment{Accepted for Publication in ApJ: November 15, 2010}

\title{
Empirical Constraints on Turbulence in Protoplanetary Accretion Disks
}

\author{A. Meredith Hughes\altaffilmark{1},
David J. Wilner\altaffilmark{1},
Sean M. Andrews\altaffilmark{1},
Chunhua Qi\altaffilmark{1},
Michiel R. Hogerheijde\altaffilmark{2}
}
\altaffiltext{1}{Harvard-Smithsonian Center for Astrophysics, 60 Garden Street, Cambridge, MA 02138; mhughes, dwilner, cqi, sandrews$@$cfa.harvard.edu}
\altaffiltext{2}{Leiden Observatory, Leiden University, P.O. Box 9513, 2300 RA,
  Leiden, The Netherlands; michiel$@$strw.leidenuniv.nl}

\begin{abstract}
We present arcsecond-scale Submillimeter Array observations of the
CO(3-2) line emission from the disks around the young stars HD~163296 and
TW~Hya at a spectral resolution of 44\,m\,s$^{-1}$.  These observations probe
below the $\sim$100\,m\,s$^{-1}$ turbulent linewidth inferred from
lower-resolution observations, and allow us to place constraints on the
turbulent linewidth in the disk atmospheres.  We reproduce the observed CO(3-2)
emission using two physical models of disk structure: (1) a power-law
temperature distribution with a tapered density distribution following a simple
functional form for an evolving accretion disk, and (2) the radiative transfer
models developed by D'Alessio et al. that can reproduce the dust emission probed
by the spectral energy distribution.  Both types of models yield a low upper
limit on the turbulent linewidth (Doppler b-parameter) in the TW~Hya system
($\lesssim$40\,m\,s$^{-1}$), and a tentative ($3\sigma$) detection of a
$\sim$300\,m\,s$^{-1}$ turbulent linewidth in the upper layers of the
HD~163296 disk.  These correspond to roughly $\leq$10\% and 40\% of
the sound speed at size scales commensurate with the resolution of the data.
The derived linewidths imply a turbulent viscosity coefficient, $\alpha$, of
order 0.01 and provide observational support for theoretical predictions
of subsonic turbulence in protoplanetary accretion disks.
\end{abstract}
\keywords{circumstellar matter --- planetary systems: protoplanetary disks ---
stars: individual (HD~163296, TW~Hydrae)}

\section{Introduction}

The accretion disks around young stars provide the raw material
and physical conditions for the planet formation process.  The viscous
transport of angular momentum drives the evolution of protoplanetary
disks \citep{lyn74,har98}, determining when, where, and
how much material is available for planet formation.  An understanding of the
physical mechanisms behind the viscous transport process is therefore
central to constraining planet formation theory.  The source of viscosity is
uncertain, since molecular viscosity implies a disk evolution timescale far
longer than the observed 1-10\,Myr.  The classic result from \citet{sha73}
sets up a framework for describing the action of turbulent viscosity in 
accretion disks, which can to account for disk evolution on the appropriate 
timescales.  However, while turbulence is commonly invoked as the source of 
viscosity in disks, its physical origin, magnitude, and spatial distribution 
are not well constrained.

The mechanism most commonly invoked as the source of this turbulence
in disks around young stars is the magnetorotational instability
(MRI), in which magnetic interactions between fluid elements in the disk
couple with an outwardly decreasing velocity field to produce torques that
transfer angular momentum from the inner disk outward \citep{bal91,bal98}.
Models of disk structure indicate that the conditions for the MRI are likely
satisfied over much of the extent of a typical circumstellar disk, with the
possible exception of an annular dead zone \citep[][and references
therein]{gam05}.  MRI turbulence has also been invoked to address a wide array
of problems in planet formation theory.  For example, it has been proposed to
regulate the settling of dust particles \citep[e.g.,][]{cie07}, to explain
mixing in meteoritic composition \citep[e.g.,][]{bos04}, to form planetesimals
\citep[e.g.,][]{joh07}, and to slow planet migration \citep[e.g.,][]{nel03}.
Measurements that constrain the magnitude and physical origin of disk
turbulence therefore promise to provide important insight into the physics
of planet formation on a variety of physical and temporal scales.

The only directly observable manifestation of turbulence is the non-thermal
broadening of spectral lines.  To date, no lines have been detected in disks
that would allow an independent determination of temperature and non-thermal
broadening, similar to NH$_3$ in molecular cloud cores \citep[e.g.,][and 
references therein]{ho83}. 
Previous interferometric observations of molecular line emission from several 
disks show gas in Keplerian rotation around the star with inferred subsonic 
turbulent velocity widths, close to the scale of the spectral resolution of
\citep[$\sim$200\,m\,s$^{-1}$, e.g.][]{pie07}.  Spectroscopic observations of
infrared CO overtone bandhead emission originating from smaller disk radii
indicate larger, approximately transonic, local line broadening that may be
associated with turbulence \citep{car04}, although those
observations of optically thick lines can only probe far above the midplane.
In combination with the millimeter data, this may indicate variations of
turbulent velocity with radius.  However, this interpretation is uncertain:
it is important to exercise caution when deriving information about velocity
fluctuations on scales smaller than the spectral resolution of the data
\citep[as noted by, e.g., ][]{pie07}.  The advent of a high spectral
resolution mode of the Submillimeter Array (SMA) correlator, capable of
resolving well below the $\sim$200\,m\,s$^{-1}$ linewidths in the low-J
transitions of CO derived from lower-resolution observations, permits access
to turbulent linewidth measurements in the cold, outer regions of molecular
gas disks around young stars.

In this paper, we conduct high spectral resolution (44\,m\,s$^{-1}$)
observations of the CO(3-2) emission from the disks around two nearby young
stars, HD~163296 and TW~Hya.  These systems were selected on the basis of
their bright CO(3-2) line emission \citep[e.g.,][]{kas97,den05}, to ensure adequate
sensitivity for high-resolution spectroscopy.  They are also particularly
well-studied using spatially-resolved observations at millimeter wavelengths,
so that excellent models of the temperature and density structure of the gas
and dust disks are already available \citep{cal02,ise07,ise09,
hug08,qi04,qi06,qi08}.  Both exhibit CO(3-2) emission that is consistent with
Keplerian rotation about the central star, and neither suffers from
significant cloud contamination.  TW~Hya is a K7 star with an age of
$\sim$10\,Myr \citep{web99}, located at a distance of only $51\pm4$\,pc
\citep{mam05}.  It hosts a nearly face-on ``transition'' disk, with an
optically thin inner cavity of radius $\sim$4\,AU indicated by the SED
\citep{cal02} and interferometrically resolved at wavelengths of 7\,mm
\citep{hug07} and 10\,$\mu$m \citep{rat07}.  The low spectral resolution
CO(3-2) line emission from the disk around TW~Hya was modeled with a
50\,m\,s$^{-1}$ turbulent linewidth by \citep{qi04}.  HD~163296 is a
Herbig Ae star with a mass of 2.3\,M$_{\sun}$, located at a distance of
122\,pc \citep{anc98}.  Its massive, gas-rich disk extends to at least 500\,AU
\citep{gra00} and is viewed at an intermediate inclination angle of
45$^\circ$ \citep{ise07}.

We describe the high spectral resolution SMA observations of the CO(3-2) line
emission from TW~Hya and HD~163296 in Section~\ref{sec:hires_obs} and present
the results in Section~\ref{sec:hires_results} (with full channel maps
provided in Appendix~\ref{sec:hires_appendix}).  In
Section~\ref{sec:hires_analysis} we outline the standard procedures that
we use to model the temperature, density, and velocity structure of the
disk, including the fixed parameters and assumptions about how the turbulent
linewidth is spatially distributed.  Section~\ref{sec:hires_best}
presents the best-fit models, and the degeneracies between parameters are 
discussed in Section~\ref{sec:degen}.  We compare our results
to theoretical predictions of the magnitude and spatial distribution of
turbulence in Section~\ref{sec:hires_discussion}, and describe the implications
for planet formation.  A summary is provided in Section~\ref{sec:hires_summary}.

\section{Observations}
\label{sec:hires_obs}

The SMA observations of TW Hya took place on 2008 March 2 in the compact
configuration, with baseline lengths of 16-77\,m, and on 2008 February 20
during the move from compact to extended configuration, with baseline lengths
of 16-182\,m.  The weather was good both nights, with stable atmospheric
phases.  Precipitable water vapor levels were extremely low on February 20,
with 225\,GHz atmospheric opacities less than 0.05 throughout the night, while
the March 2 levels were somewhat higher, rising smoothly from 0.08 to 0.11.
In order to calibrate the atmospheric and instrumental gain variations,
observations of TW Hya were interleaved with the nearby quasar J1037-295.  To
test the efficacy of phase transfer, observations of 3c279 were also included
in the observing loop.  Flux calibration was carried out using observations of
Callisto; the derived fluxes of 3c111 were 0.76 and 0.78\,Jy on February 20
and March 2, respectively.

The observations of HD 163296 were carried out in the compact-north
configuration on 2009 May 6, with baseline lengths of 16 to 139\,m, and in
the extended configuration on 2009 August 23, with baseline lengths of
44 to 226\,m.  Atmospheric phases were stable on both nights, and the
225\,GHz opacities were 0.05 on May 6 and 0.10 on August 23.  The observing
loop included J1733-130 for gain calibration and J1924-292 for testing the
phase transfer.  Callisto again served as the flux calibrator, yielding
derived fluxes of 1.17 and 1.30\,Jy for 1733-130 on May 6 and August 23,
respectively.

For all observations, the correlator was configured to divide a single
104\,MHz-wide chunk of the correlator into 2048 channels.  This high-resolution
chunk was centered on the 345.796\,GHz frequency of the CO(3-2) line, yielding
a spectral resolution of 44.1\,m\,s$^{-1}$ across the line.  Because this
used up a large portion of the available correlator capacity, only 1.3\,GHz
of the 2\,GHz bandwidth in each sideband was available for continuum
observations.  The bandpass response was calibrated using extended
observations of 3c273, 3c279, and Saturn for the TW Hya tracks and 3c454.3,
Callisto, and 1924-292 for the HD 163296 tracks.  Since the sidebands are
separated by 10\,GHz and the CO(3-2) line was located in the upper sideband
for the HD 163296 observations and in the lower sideband for the TW Hya
observations, the continuum observations are at frequencies of 340\,GHz for
HD 163296 and 350\,GHz for TW Hya.

Routine calibration tasks were carried out using the MIR software
package\footnote{See http://cfa-www.harvard.edu/$\sim$cqi/mircook.html.}, and
imaging and deconvolution were accomplished with MIRIAD.  The observational
parameters, including the rms noise for both the line and continuum data, are
given in Table~\ref{tab:obs}.  Note that due to weather the compact
observations of HD~163296 were substantially more sensitive than the extended
data and so the combined data set is dominated by the compact data; the
reverse is true for TW~Hya.

\begin{table*}
\caption{Observational Parameters$^a$}
\resizebox{\textwidth}{!}{
\begin{tabular}{lcccccc}
\hline
\hline
 & \multicolumn{3}{c}{HD~163296} & \multicolumn{3}{c}{TW~Hya} \\
\cline{2-4} \cline{5-7}
 & Compact-N & Extended & C+E & Compact & Extended & C+E \\
Parameter & 2009 May 6 & 2009 August 23 & & 2008 March 2 & 2008 February 20 & \\
\hline
 & \multicolumn{6}{c}{CO(3-2) Line} \\
\hline
Beam Size (FWHM) & 2\farcs1$\times$1\farcs4 & 0\farcs9$\times$0\farcs7 & 1\farcs7$\times$1\farcs3 & 1\farcs0$\times$0\farcs8 & 1\farcs0$\times$0\farcs7 & 1\farcs0$\times$0\farcs8 \\
~~~P.A. & 50$^\circ$ & 8$^\circ$ & 47$^\circ$ & 5$^\circ$ & -17$^\circ$ & -16$^\circ$ \\
RMS Noise (Jy\,beam$^{-1}$) & 0.51 & 0.97 & 0.49 & 0.35 & 0.52 & 0.40 \\
Peak Flux Density (Jy\,beam$^{-1}$) & 8.9 & 3.1 & 8.1 & 4.8 & 4.0 & 4.8 \\
Integrated Flux$^b$ (Jy\,km\,s$^{-1}$) & 76 & 14 & 76 & 19 & 4.8 & 24 \\
\hline
 & \multicolumn{3}{c}{340\,GHz Continuum} & \multicolumn{3}{c}{350\,GHz Continuum} \\
\hline
Beam Size (FWHM) & 2\farcs1$\times$1\farcs4  & 0\farcs9$\times$0\farcs7 & 1\farcs7$\times$1\farcs3 & 1\farcs0$\times$0\farcs9 & 1\farcs0$\times$0\farcs7 & 1\farcs0$\times$0\farcs8\\
~~~P.A. & 52$^\circ$ & 9$^\circ$ & 47$^\circ$ & 8$^\circ$ & -21$^\circ$ & -21$^\circ$ \\
RMS Noise (mJy\,beam$^{-1}$) & 7.0 & 10 & 7.0 & 16 & 10 & 8.5 \\
Peak Flux Density (Jy\,beam$^{-1}$) & 1.14 & 0.6 & 1.05 & 1.21 & 0.47 & 0.51 \\
Integrated Flux$^c$ (Jy) & 1.78 & 1.72 & 1.75 & 1.67 & 1.49 & 1.57 \\
\hline
\end{tabular}
}
\tablenotetext{a}{All quoted values assume natural weighting.}
\tablenotetext{b}{The integrated line flux is calculated by integrating the
zeroth moment map inside the 3$\sigma$ brightness contours using the MIRIAD
task \texttt{cgcurs}.}
\tablenotetext{c}{The integrated continuum flux is calculated using the MIRIAD task \texttt{uvfit}, assuming an elliptical Gaussian brightness profile.}
\label{tab:obs}
\end{table*}

\section{Results}
\label{sec:hires_results}

\begin{figure*}
\epsscale{0.9}
\plotone{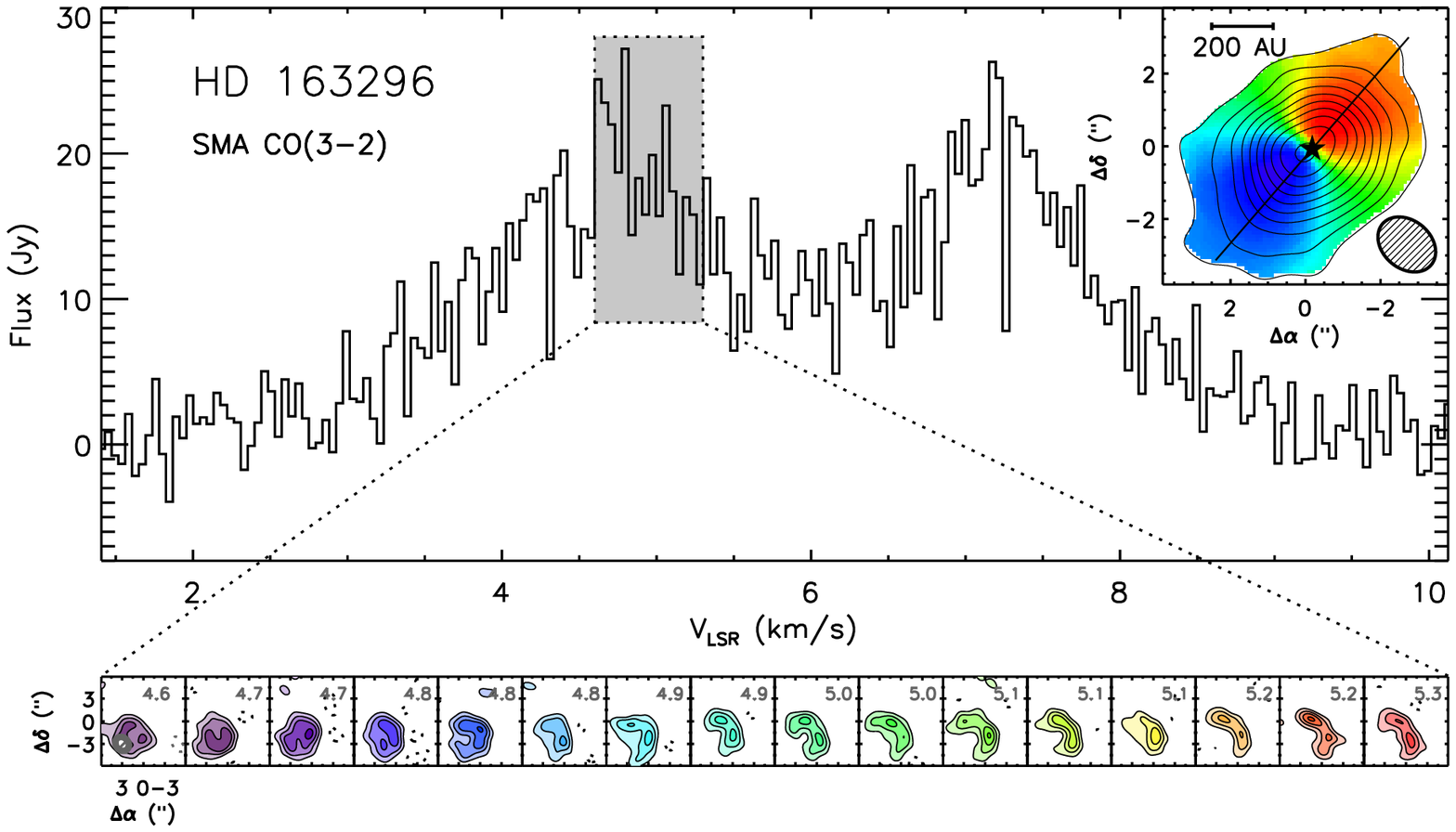}
\figcaption{CO(3-2) emission from the disk around HD~163296 observed with the
SMA at a spectral resolution of 44\,m\,s$^{-1}$.  Top plot shows
the line profile, summed within a 6 arcsec box using the MIRIAD task 
\texttt{imspec} (neglecting emission with absolute values between 
$\pm$2$\sigma$).  Channel maps across the bottom show the segment of the line 
indicated by the shaded gray box at its full spatial and spectral resolution, 
imaged with a 1\farcs0 taper to bring out the large-scale emission (complete
channel maps are provided in Appendix~\ref{sec:hires_appendix}).  LSR
velocity is indicated by the numbers in the upper right of each channel.
Contours are [3,6,9,...]$\times$0.55\,Jy\,beam$^{-1}$ (the rms noise).
Inset in the upper right corner is a zeroth (contours) and first (colors) 
moment map of the CO(3-2) line emission.  The 2\farcs0$\times$1\farcs7 beam 
is indicated in the lower right of the inset.  Note that while the colors in 
the channel and moment maps both represent LSR velocity (blue is low; red is 
high), the scales are different for the two representations: the moment map 
contains the full line data, while the channel maps span only a subset of 
the line.
\label{fig:CO_hd}}
\end{figure*}

\begin{figure*}[t]
\epsscale{1.0}
\plotone{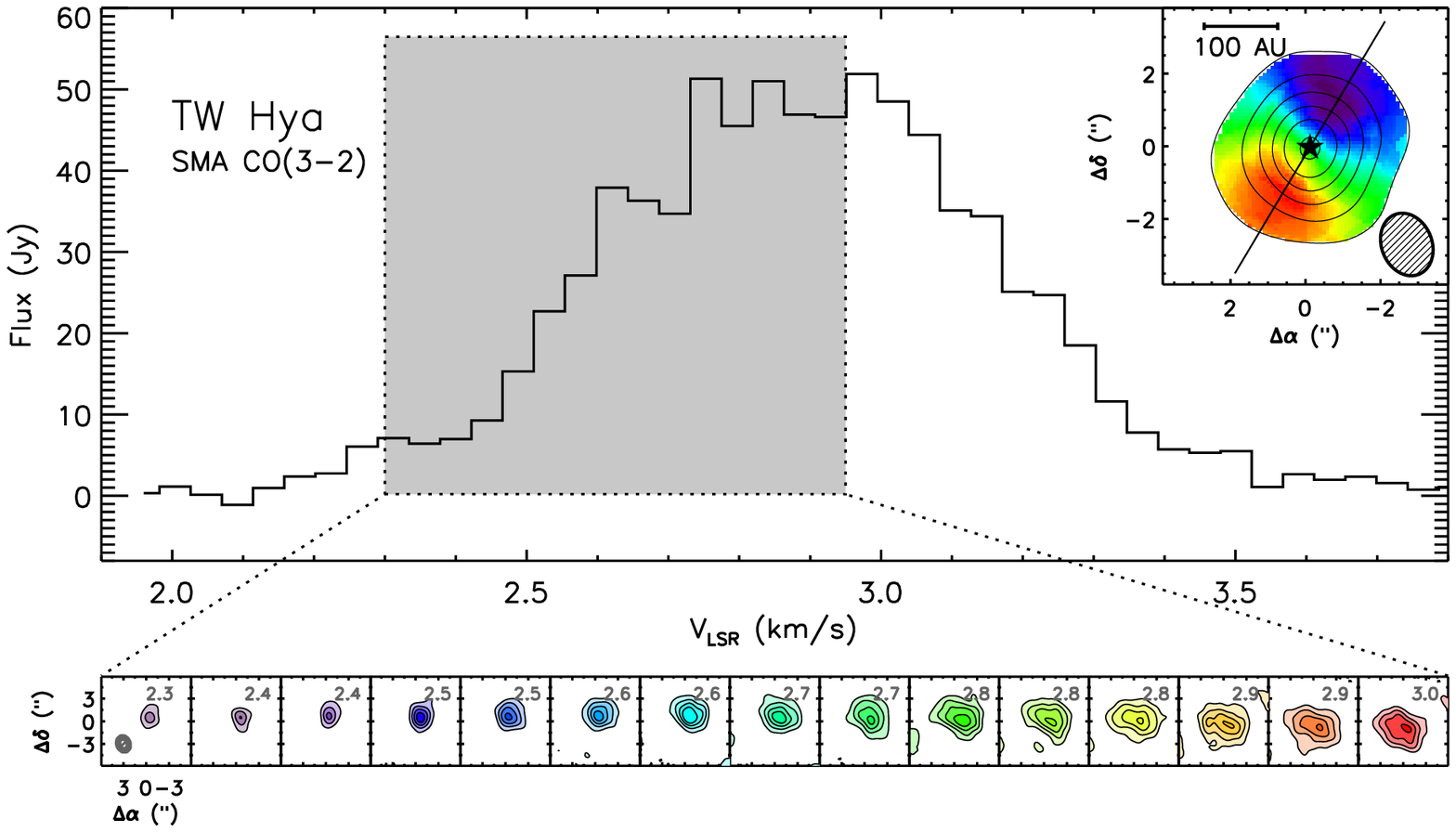}
\figcaption{Same as Figure~\ref{fig:CO_hd} but for TW Hya.  The channel maps
were imaged with a 1\farcs2 Gaussian taper to emphasize the emission on larger
scales, and the contours are [4,8,12,...]$\times$0.55\,Jy\,beam$^{-1}$ (the
rms noise).  For the full set of channel maps, see
Appendix~\ref{sec:hires_appendix}.
\label{fig:CO_twh}}
\end{figure*}

We detect CO(3-2) emission at 44\,m\,s$^{-1}$ resolution from both TW Hya
and HD~163296 in the compact and extended configurations.
Figures~\ref{fig:CO_hd} and \ref{fig:CO_twh} present the line emission
from HD 163296 and TW Hya, respectively.  The upper left panel shows the full
line profile summed within a 6 arcsec square box (neglecting emission within the
range $\pm$2$\sigma$), with emission detected across $\sim$50 channels for
TW~Hya and $\sim$200 for HD~163296.  Beneath the line profiles are the
spatially resolved channel maps for a subset of the data, indicated by the
gray box around the line peak.  In the upper right are the zeroth (contours)
and first (colors) moment maps: these are the velocity-integrated intensity and
intensity-weighted velocity of the emission, respectively.  The line emission
is regular, symmetric, and consistent with material in Keplerian rotation
around the central star viewed at an inclination to our line of sight.

The peak and integrated fluxes for each of the four tracks and the combined
data sets are listed in Table~\ref{tab:obs}.  Appendix~\ref{sec:hires_appendix}
presents the full channel maps of the combined (compact and extended
configuration) data set for each source.

%

\section{Analysis}
\label{sec:hires_analysis}

In order to constrain the turbulent linewidth in the disks around TW~Hya and
HD~163296, we fit models of the temperature, density, and velocity structure
to the high spectral resolution CO(3-2) line data.  For the initial modeling
effort presented here, we use two well-tested physical models of disk
structure: (1) power-law models of the disk temperature structure combined
with tapered surface density profiles corresponding to the functional form
predicted for a simple viscous accretion disk \citep{lyn74,har98}, and (2)
the 1+1D radiative transfer models developed by D'Alessio et al. to
reproduce the dust emission represented in the SEDs of young systems.  These
models are described in more detail in Section~\ref{sec:model_description}.

We use these two classes of models because they are well established and have
been successful in describing the observed structure of circumstellar disks
across a wide range of wavelengths, particularly in the submillimeter
\citep[see, e.g.,][]{cal02,cal05,and09a}.  However, each class of models has
limitations.  The similarity solution models have a large number of free
parameters, some with significant degeneracies \citep[see discussion in][]{
and09a}.  By fitting only the CO(3-2) emission, these models
also neglect potential information provided by dust emission, including
stronger constraints on the disk density.  However, the neglect of dust
emission avoids complications due to heating processes and chemistry that
affect gas differently than dust.  The D'Alessio et al. models of dust emission
include only stellar irradiation and viscous dissipation as heating sources,
and do not take into account the additional heating processes that may affect
molecular line strengths in the upper layers of circumstellar disks
\citep{qi06}.  While the constraints from the dust continuum reduce the number
of free parameters in this class of models, they also have the disadvantage
of an unrealistic treatment of the density structure at the disk outer edge:
since they are simply truncated at a particular outer radius, they are not
capable of simultaneously reproducing the extent of gas and dust emission in
these systems \citep{hug08}.  The similarity solution models are vertically 
isothermal, which is an unrealistic assumption.  With only one molecular line 
included in the model, this limitation will not affect the results of the 
study presented here, although caution should be exercised when applying the 
best-fit model parameters to other lines.

The primary reason for using the two types of models, however, is that they
differ substantially in their treatment of the disk temperature structures.
For the D'Alessio et al. models, the temperature structure is fixed by the dust
continuum.  The similarity solution models, by contrast, allow the temperature
to vary to best match the data.  There are a few independent constraints on
temperature: it should increase with height above the midplane, due to surface
heating by the star and low viscous heating in the midplane, and the dust will
generally not be hotter than the gas, since the gas is subject to additional
heating processes beyond the stellar irradiation that determines dust
temperature.  The temperature structure in the disk is the single factor 
most closely tied to the derived value of the turbulent linewidth (see 
discussion in Section \ref{sec:degen}), which will be model-dependent.  We 
therefore fit both classes of models to the data, in order to compare the 
model-dependent conclusions about turbulent linewidth for two distinct types 
of models with very different treatments of gas temperature.  The spatial 
dynamic range of the data is insufficient to investigate radial variations in 
turbulent linewidth.  We therefore assume a global value, $\xi$, that will 
apply to size scales commensurate with the spatial resolution of the data.

\subsection{Description of Models}
\label{sec:model_description}

\subsubsection{D'Alessio et al. Models}

The D'Alessio et al. models are described in detail in 
\citet{dal98,dal99,dal01,dal06}.  Here we provide a general outline of the 
model properties and discuss the particular models used in this paper.

The D'Alessio et al. models were developed to reproduce the unresolved SEDs
arising from warm dust orbiting young stars, although they have also been
demonstrated to be successful at reproducing spatially resolved dust
continuum emission at millimeter wavelengths \citep[see, e.g.,][]{cal02,hug07,
hug09} as well as spatially-resolved molecular line emission \citep[see,
e.g.,][]{qi04,qi06}.  The models include heating from the central star and
viscous dissipation within the disk, although they tend to be dominated by
stellar irradiation.  The structure is solved iteratively to provide
consistency between the irradiation heating and the vertical structure.
The mass accretion rate is assumed to be constant throughout the disk.  The
assumed dust properties are described by \citet{cal02}, and the model includes
provisions for changing dust properties, dust growth, and settling.  We allow
the outer radius of the model to vary to best reproduce the extent of the
molecular line observations.

We use the structure model for TW~Hya that was developed by \citet{cal02} and
successfully compared to molecular line emission by \citet{qi04,qi06}.  For
HD~163296, we use a comparable model that reproduces the spatially unresolved
SED and is designed to reproduce the integrated line strengths of several
CO transitions as well as other molecules (Qi et al., in prep).

Since the D'Alessio et al. models were developed primarily to reproduce the
dust emission from the SED, we are required to fit several parameters to
match the observed CO(3-2) emission using the SED-based models.  We fit the
structural parameters \{$R_D$, $X_\mathrm{CO}$\} (the disk outer radius and
CO abundance, respectively), the geometrical parameters \{$i$, PA\} (the disk
inclination and position angle), and the turbulent linewidth, \{$\xi$\}.

\subsubsection{Viscous Disk Similarity Solutions}

We also fit the observations using a power-law temperature distribution and
surface density profile that follows the class of similarity solutions for
evolving viscous accretion disks described by \citet{lyn74} and \citet{har98}.
This particular method of parameterizing circumstellar disk structure has a
long history of success in reproducing observational diagnostics, although
with limitations.  Theoretical predictions of the power-law dependence of
temperature for accretion disks around young stars were first made by
\citet{ada86}, and power-law parameterizations of temperature and surface
density have been used by many studies since then \citep[e.g.][]{bec90,bec91,
mun93,dut94,lay94,and07}.  The similarity solutions are equivalent to a
power-law surface density description in the inner disk, but with an
exponentially tapered outer edge, which was shown by \citet{hug08} to better
reproduce the extent of gas and dust emission than traditional power-law
descriptions with abruptly truncated outer edges.  Recent high spatial
resolution studies have used this class of models to reproduce successfully
the extent of gas and dust emission in circumstellar disks from several nearby
star-forming regions \citep[e.g.][]{and09a,ise09}.

The temperatures and surface densities of these models are parameterized as
follows:
\begin{eqnarray}
T(R) = T_{100} \left( \frac{R}{100 \mathrm{AU}} \right) ^{-q} \\
\Sigma(R) = \frac{c_1}{R^\gamma} \exp{\left[-\left(\frac{R}{R_c}\right)^{2-\gamma} \right]}
\end{eqnarray}
where $R$ is the radial distance from the star in AU, $T_{100}$ is the
temperature indicated by the CO(3-2) line at 100\,AU from the star, $q$
describes how the temperature decreases with distance from the star, $c_1$
is a constant describing the surface density normalization, $R_c$ is a
constant related to the radial scale on which the exponential taper decreases
the disk density, and $\gamma$ describes how surface density falls with radius
in the inner disk regions \citep[comparable to the parameter $p$ in typical
power-law descriptions of surface density; see e.g.][for a description of
the power-law model parameters]{dut94}.  Because the high optical depth of
the CO(3-2) line is a poor tracer of the radial dependence of $\Sigma$, we
fix $\gamma$ at a value of 1 for both systems, which is consistent with
theoretical predictions for a constant-$\alpha$ accretion disk \citep{har98},
as well as observations of young disks in Ophiuchus \citep{and09a} and
previous studies of the continuum emission from these systems \citep{hug08}.
We therefore fit the high spectral resolution CO(3-2) observations using four
structural parameters, \{$T_{100}$, $q$, $c_1$, $R_c$\}, two geometric
parameters, \{$i$, PA\}, and the turbulent linewidth, \{$\xi$\}.

\subsection{Modeling Procedure}
\label{sec:procedure}

We assume that the disks have a Keplerian velocity field, using stellar
masses and distances from the literature to model the rotation pattern
\citep[0.6\,M$_{\sun}$ at 51\,pc for TW~Hya and 2.3\,M$_{\sun}$ at 122\,pc for
HD~163296; see][]{cal02,web99,mam05,anc98}.  Gas and dust are assumed to be
well-mixed in both models; the gas-to-dust mass ratio is fixed at 100 while
the CO abundance is allowed to vary in the D'Alessio et al. models in order to
reproduce the CO emission while maintaining consistency with the dust continuum
emission from which the model was derived.  Since we don't include continuum
emission in the fits, we fix the CO abundance at $10^{-4}$ for the similarity
solution models because there is no constraint on the relative content of gas
and dust.  In regions of the disk where the temperature drops below 20\,K, the
CO abundance is reduced by a factor of $10^{4}$ to simulate the effects of
freeze-out onto dust grains.  We assume a global, spatially uniform value of
the turbulent linewidth, which is implemented as an addition to the thermal
linewidth.

\begin{figure}[t]
\epsscale{0.8}
\plotone{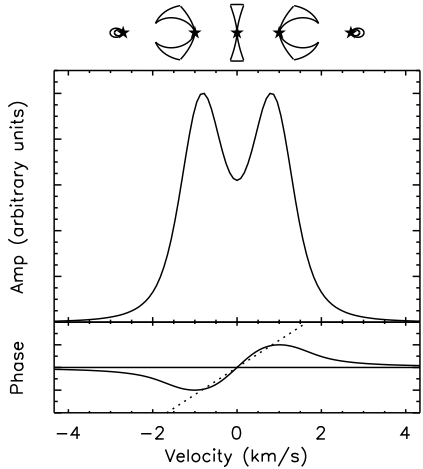}
\figcaption{
Cartoon showing the amplitude and phase of a spectral line as a function of 
velocity and how they correspond to the spatial distribution of flux in the
disk.  The line wings originate from small areas of fast-moving material close 
to the star, resulting in low amplitude and small phase offsets.  The line
peaks arise from a large surface area of intermediate-velocity material with
a large spatial offset from the star, resulting in high amplitudes and large
visibility phases. The line center is symmetric and therefore has zero phase.  
We determined the systemic velocity by assuming an axisymmetric flux
distribution and fitting a line to the central channels of the visibility 
phases (similar to the dotted line in the figure).  The systemic 
velocity is the point at which the asymmetry changes sign and the line crosses 
zero.  
\label{fig:toy_phase}}
\end{figure}

In order to compare the models with the data, the systemic velocity, or central
velocity of the line, must be determined.  We calculated the visibility phases
for the spatially integrated CO(3-2) emission and fit a line to the central
few channels (45 for HD~163296; 12 for TW~Hya) to determine the systemic
velocity.  Figure~\ref{fig:toy_phase} presents a cartoon of this method, which 
is largely model-independent, requiring only the assumption that the flux 
distribution is axisymmetric.  The visibility phases encode information about 
the spatial symmetry of the line emission in each channel: for a Keplerian 
disk, it is most symmetric, and therefore the phase is zero, at line center.  
The channel maps are most asymmetric about disk center near the line peaks, 
with opposite spatial offsets at the two line peaks.  The phase therefore 
reverses sign between one peak and the other, with an approximately linear 
relationship near the line center.  The linear fit therefore uses the 
integrated emission from the central few channels to pinpoint the precise 
location where the phase is zero and the line is most symmetric, i.e., at the 
systemic velocity (it should be noted that if there were large-scale 
asymmetries in the CO data this method would not produce a reliable systemic 
velocity, but we see no evidence of such asymmetries in the data).  We derive 
a systemic velocity of 5.79\,km\,s$^{-1}$ for HD~163296 and 2.86\,km\,s$^{-1}$ 
for TW Hya.  In order to generate model emission with the appropriate velocity 
offset, we create a model image at higher spectral resolution than the data, 
and then re-bin it at the appropriate velocity sampling using the MIRIAD task 
\texttt{regrid}.

Each disk model specifies a particular density, temperature, and velocity
structure.  We use the Monte-Carlo radiative transfer code RATRAN
\citep{hog00} to calculate equilibrium populations for each rotational level
of the CO molecule and generate a sky-projected image of the CO(3-2) line
emission at a given viewing geometry \{$i$,PA\} for each model.  We compare
these simulated models directly to the data in the Fourier domain.  In order
to sample the model images at the appropriate spatial frequencies for
comparison with the SMA data, we use the MIRIAD task \texttt{uvmodel}.  We
then compute the $\chi^2$ statistic for each model compared with the data
using the real and imaginary simulated visibilities.  Due to the high
computational intensity of the molecular line radiative transfer, it is
prohibitively time-intensive to generate very large and well-sampled grids
of models for the $\chi^2$ comparison.  Instead, we move from coarsely-sampled
grids that cover large regions of parameter space to progressively more
refined (but still small) grids to avoid landing at a local minimum.  However,
this has the result that the degeneracies of the parameter space are poorly
characterized.  A discussion of these degeneracies is included in
Section~\ref{sec:degen} below.

\subsection{Best-fit Models}
\label{sec:hires_best}

\begin{figure*}
\epsscale{1.0}
\plottwo{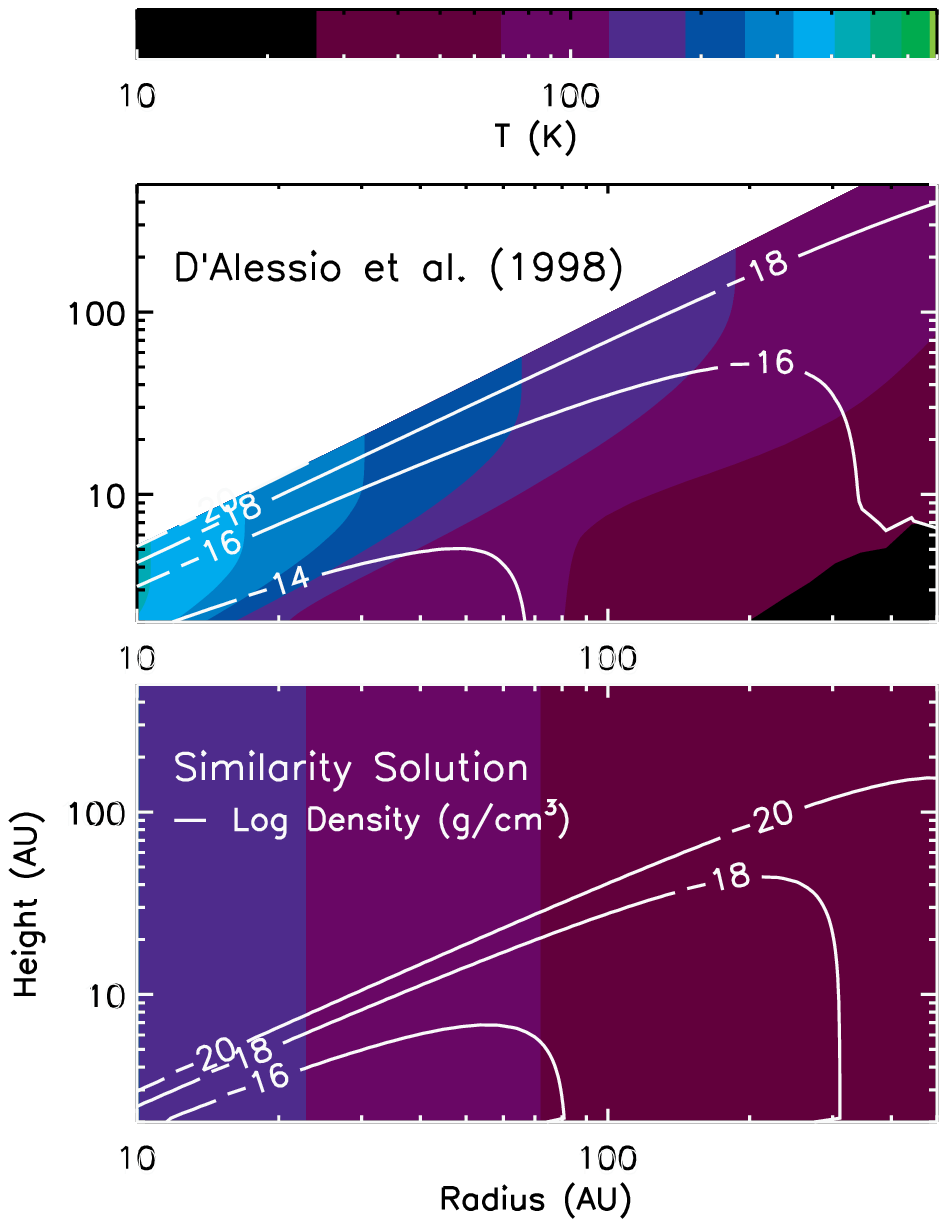}{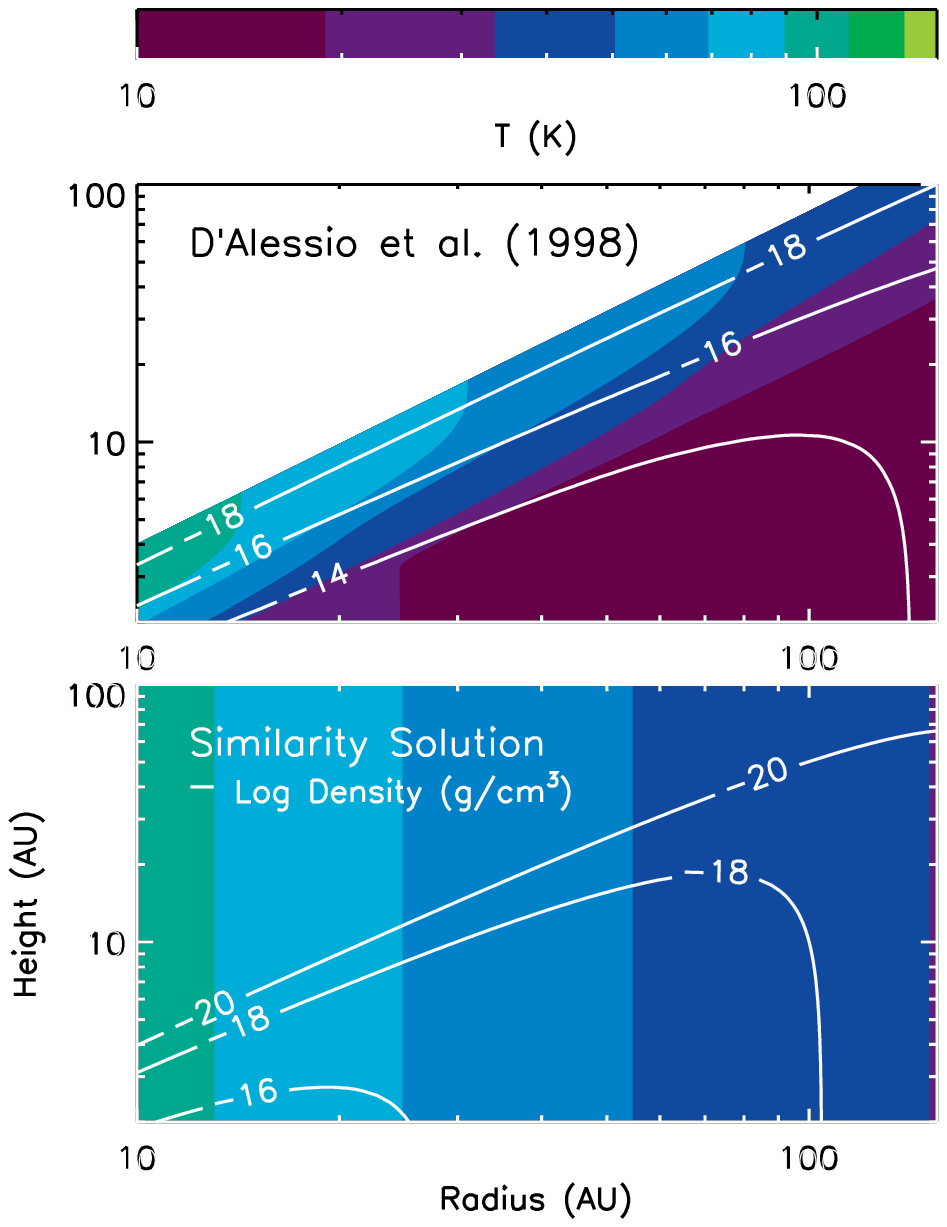}
\figcaption{
Comparison between the temperature and density structures of the similarity
solution and D'Alessio et al. models for HD 163296 ({\it left}) and TW Hya
({\it right}).  The bar across the top shows the temperature scale represented
by the colors, while the white contours represent density, marked with the
base 10 logarithm of the total (gas+dust) mass density in units of
g\,cm$^{-3}$.  Note that while the D'Alessio et al. model of HD~163296 appears
to be two orders of magnitude more dense than the similarity solution model,
the CO abundance is more than two orders of magnitude lower, making the CO mass
densities comparable. The CO abundance of the TW Hya model is one order of
magnitude lower than for the similarity solution.  The abscissae of the plots
are scaled to match the radial extent of the power-law disk model (excising
the central 10\,AU, which are not accessible with the data), although the
similarity solution models will extend farther.
\label{fig:temp_dens}}
\end{figure*}

The best-fit parameters for both types of models are presented in
Table~\ref{tab:hires_best}.  Their temperature and density structures are
plotted in Figure~\ref{fig:temp_dens}.  Note that the midplane temperatures
for the similarity solution models are likely much lower than indicated; the
power-law temperature representation parameterizes only the radial dependence
of temperature in the upper disk layers from which the optically thick CO(3-2)
emission arises.

\begin{figure*}
\epsscale{1.0}
\plotone{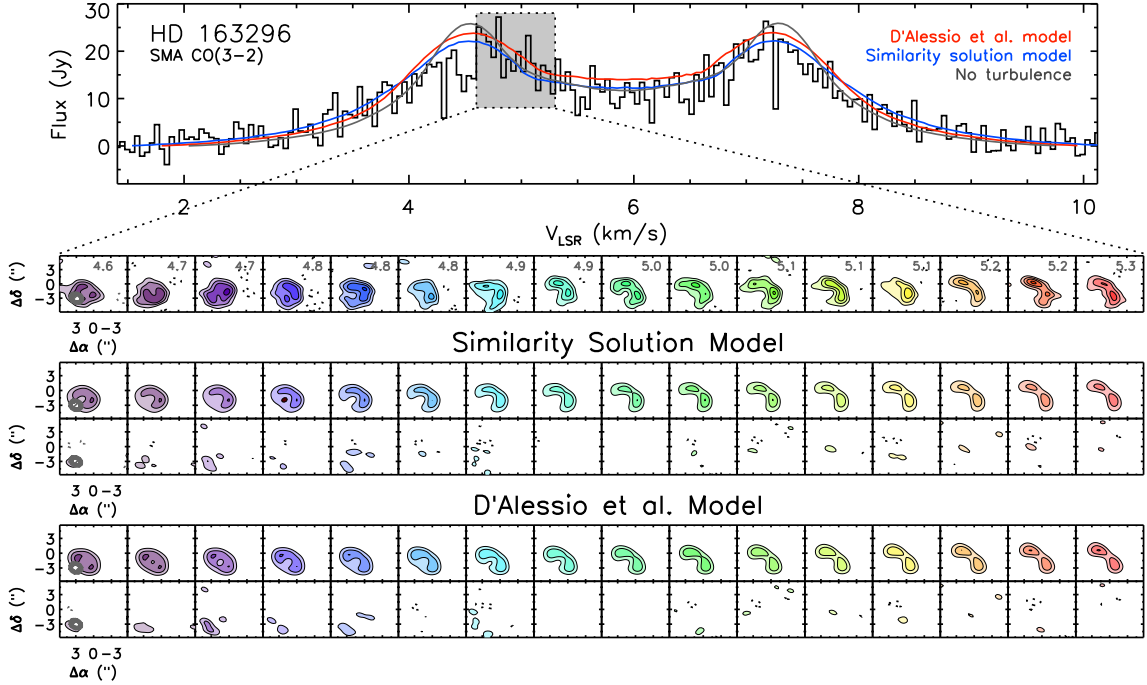}
\figcaption{
Comparison of CO(3-2) emission from HD~163296 between the data and best-fit
models for a subset of the data.  The top row shows the same subset of
channels as in Figure~\ref{fig:CO_hd}.  The central set of channel maps shows
the corresponding channels of the best-fit similarity solution model
and the residuals (subtracted in the visibility domain).  The bottom set
of channel maps shows the best-fit D'Alessio et al. model and residuals.  
Contour levels, beam sizes, and imaging parameters are identical to those in
Figure~\ref{fig:CO_hd}.  An additional line profile for the best-fit similarity
solution without turbulence is overplotted in gray.
\label{fig:model_chmap_hd}}
\end{figure*}

\begin{figure*}[t]
\epsscale{1.0}
\plotone{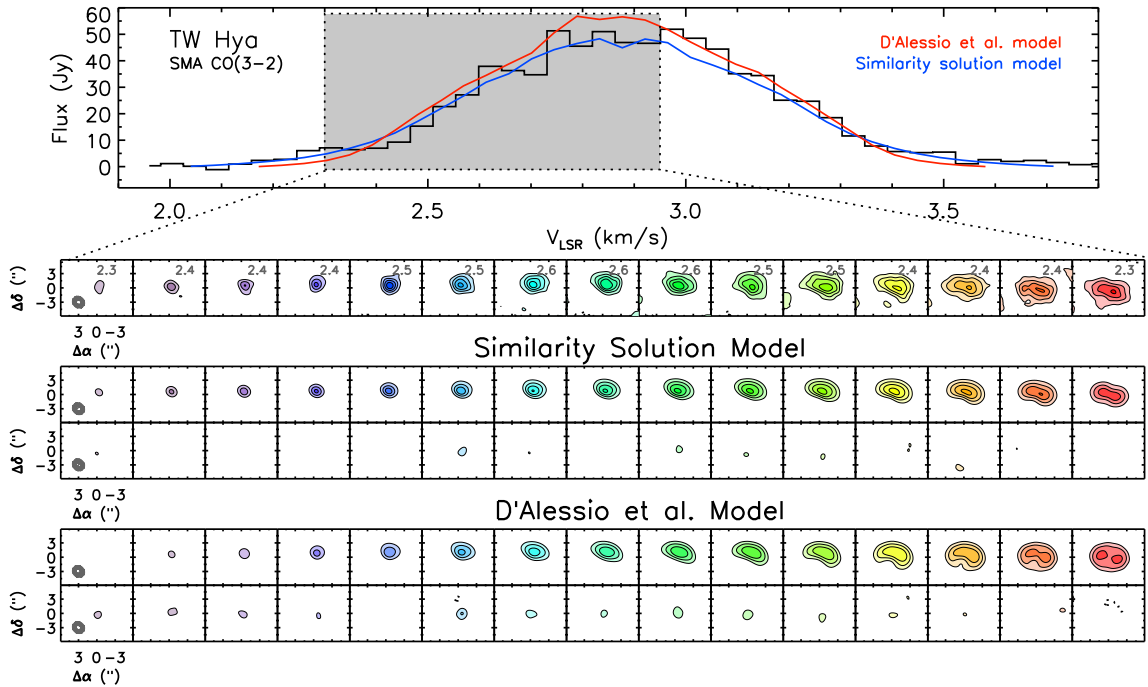}
\figcaption{
Same as Figure~\ref{fig:model_chmap_hd} but for TW~Hya.  The channels,
contour levels, and imaging parameters are identical to those in
Figure~\ref{fig:CO_twh}.
\label{fig:model_chmap_twh}}
\end{figure*}

The $\chi^2$ values for the similarity solutions are lower than for the
D'Alessio et al. models; this may be due to gradients between gas and dust
properties that influence the D'Alessio et al. model fit but not the similarity
solution models.  A sample of channel maps comparing the data with the two
classes of models are presented in Figures~\ref{fig:model_chmap_hd} and
\ref{fig:model_chmap_twh}.  From the residuals in the D'Alessio et al. model of
TW~Hya, there is evidence of a mismatch in the temperature gradient between
the model and the data: the residuals are systematically more positive near
the disk center and negative farther from the star.  For HD~163296, the
difference is more subtle: the residuals are small and apparently
spatially random (with the exception of some positive emission seen near a
velocity of 4.7\,m\,s$^{-1}$ but not near the corresponding position in the
mirror half of the line).  It should be noted that the CO abundance derived
for this source is extremely low, nearly three orders of magnitude below the
standard value of $10^{-4}$.  The reason for this is most likely an
overestimate of the temperature in the upper disk layers.  In the absence of
better information, this SED model was created with a very low turbulent
linewidth ($\sim$50\,m\,s$^{-1}$) and correspondingly little stirring of
large dust grains above the midplane.  The addition of a turbulent linewidth
comparable to the best-fit value for the CO lines would substantially reduce
settling and lower the temperature of the upper disk layers as more of the
mass is placed in large dust grains.  Such a model is under development by
Qi et al. (in prep), and can also aid in explaining the spatial distribution
of multiple CO transitions and CO isotopologue emission from this disk.

One important outcome of the modeling process is the consistency in the
measurement of turbulent linewidth in each source for the two types of models,
despite the differences in their treatment of temperature.  In both types of
models for HD~163296, the best-fit model with turbulence fits the
data better than a comparable model without turbulence at the $\sim3\sigma$
level.  If the turbulent linewidth is fixed at 0\,m\,s$^{-1}$ in
the similarity solution and the temperature allowed to vary to compensate, the
parameter $T_{100}$ must increase to 77\,K; even then, the $\chi^2$ for a
model with higher temperature and no turbulence is a poorer fit than the
best-fit model with turbulence at the $\sim3\sigma$ level (the line profile
for this model is plotted in Figure~\ref{fig:model_chmap_hd}).  The TW~Hya data
are consistent with no turbulent linewidth whatsoever.

\begin{table}
\caption{Best-Fit Model Parameters}
\begin{center}
\begin{tabular}{l cc}
\hline
Parameter & HD~163296 & TW~Hya \\
\hline
\hline
\multicolumn{3}{c}{Similarity Solution} \\
\hline
$T_{100}$ (K) & 60 & 40\\
$q$ & 0.5 & 0.4 \\
$c_1$ (cm$^{-2}$) & $1.0\times10^{12}$ & $1.0\times10^{11}$ \\
$R_c$ (AU) & 150 & 50 \\
$\xi$ (m\,s$^{-1}$) & 300 & $\lesssim 40$ \\
$i$ ($^\circ$) & 40 & 6.0 \\
PA ($^\circ$) & 131 & 155 \\
Reduced $\chi^2$ & 2.642 & 2.106 \\
\hline
\multicolumn{3}{c}{D'Alessio et al. Model} \\
\hline
$R_D$ (AU) & 525 & 155 \\
$X_\mathrm{CO}$ & $5\times10^{-7}$ & $1.5\times10^{-5}$ \\
$\xi$ (m\,s$^{-1}$) & 300 & $\lesssim 40$ \\
$i$ ($^\circ$) & 40 & 5 \\
PA ($^\circ$) & 138 & 155 \\
Reduced $\chi^2$ & 2.885 & 2.108 \\
\hline
\end{tabular}
\end{center}
\label{tab:hires_best}
\end{table}

\subsection{Parameter Degeneracies}
\label{sec:degen}

In order to better characterize our ability to measure turbulent linewidth,
it is important to understand its relationship to the other parameters.  The
interdependence between parameters {\it other} than the turbulent linewidth has
been explored at length in previous papers \citep[see, e.g., discussion of
similarity solution parameters in][]{and09a}, so here we focus on the
relationships and degeneracies specific to the turbulent linewidth.  There
are four main categories of line broadening in circumstellar disks that are relevant to our investigation: rotational, thermal, turbulent,
and optical depth.  These types of line broadening are all incorporated in
detail into the ray-tracing portion of the RATRAN radiative transfer code,
and will be handled appropriately for a given disk structure.  The goal is
to understand how to distinguish the distinct contributions of each of these
different sources of line broadening and their relationships to the parameters
of our disk structure models.

As discussed above, a detailed characterization of the multi-dimensional
parameter space is prohibitively computationally expensive.  We therefore
investigate parameter relationships by letting the two-dimensional $\chi^2$ 
values generated in Section \ref{sec:procedure} guide an investigation using 
a toy model of an optically thick spectral line profile to highlight the 
distinct contribution of each related parameter to the observable properties.

The $\chi^2$ values indicate that for the similarity solution models, the 
parameters that are most strongly degenerate with the turbulent linewidth are 
the temperature ($T_{100}$ and $q$) and inclination ($i$). This is unsurprising,
given the obvious relationship between temperature and thermal broadening and
between inclination and rotational broadening.  The optically thick CO(3-2)
line responds only weakly to variations in density, and the outer radius and
position angle of emission should intuitively be unrelated to line broadening,
hence the independence of turbulent linewidth from $c_1$, $R_c$, and PA.  For
the D'Alessio et al. models, inclination and CO abundance ($i$ and $X_{CO}$) have
the strongest relationships with turbulent linewidth.  The contribution of the
CO abundance in this case can be understood as a thermal broadening effect:
because of the vertical temperature gradient (see Figure~\ref{fig:temp_dens}),
the CO abundance controls the location of the $\tau$=1 surface and therefore
the apparent temperature of the CO(3-2) line emission.

To characterize the effects of these variables on the observable properties
of the CO(3-2) emission, we investigate their influence on a toy model of
optically thick line emission.  We assume a power-law temperature distribution
for a geometrically flat, optically thick, azimuthally symmetric circumstellar
disk.  In the Rayleigh-Jeans approximation, the brightness of the line at a
given frequency will be directly proportional to the temperature.  We
include two sources of line broadening, thermal and turbulent, implemented
by the relationship $\Delta v(r) = \sqrt{2 k_B T(r)/m + \xi^2}$, where
$\Delta v$ is the total linewidth, $\xi$ is the turbulent linewidth, and the
thermal linewidth is $\sqrt{2 k_B T/m}$ where $k_B$ is Boltzmann's constant,
$T$ is the local temperature in the disk, and $m$ is the average mass per
particle.  Rotational broadening is implicitly included in the assumed
Keplerian rotation pattern of the material, so that in polar coordinates, the
central frequency of the line as a function of position is given by
$\nu(r,\theta) = \nu_0 / c \sqrt{GM_{*}/r} \sin{i} \cos{\theta}$, where
$\nu_0$ is the line frequency and $M_{*}$ is the stellar mass.
The line profile at any point in the disk is then given by
\begin{eqnarray}
\phi(\nu,r) \propto \frac{T(r)}{\Delta v(r) \nu_0 / c} \exp \left[ \frac{-(\nu - \nu_0)^2}{(\Delta v(r) \nu_0 / c)^2} \right]
\end{eqnarray}
where $\nu_0$ is the frequency of the line center and $c$ is the speed of light.
We project the disk onto the sky according to its inclination, and integrate 
over the $r$ and $\theta$ coordinates across the extent of the elliptical disk 
to calculate the line profile as a function of frequency.
We investigate the contribution of the different sources of broadening by
varying $\xi$, $T(r)$, and $i$ to correspond to the turbulent, thermal, and
rotational broadening effects.  Here we discuss the ways in which turbulent
broadening can be distinguished from the other two effects in the context of
our toy model.

\begin{figure*}[t]
\epsscale{1.0}
\plotone{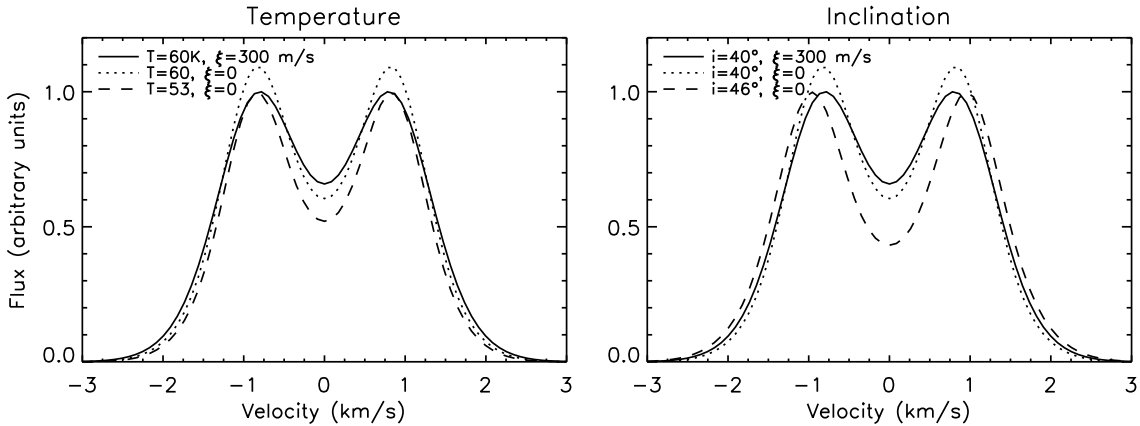}
\figcaption{Model spectral line profiles for a geometrically flat, optically
thick, azimuthally symmetric circumstellar disk (for details, see toy model
description in Section~\ref{sec:degen}).  The left panel explores the
relationship between thermal and turbulent line broadening, while the right
panel explores the relationship between turbulent broadening and rotational
broadening through changes in inclination.  The solid line in each panel is
generated using values from the best-fit similarity solution model
($T_{100}$=60\,K, $q$=0.5, and $\xi$=300\,m\,s$^{-1}$), while the dotted lines
are the same model without turbulence, and the dashed lines seek to compensate
for the lack of turbulence through the alternative line broadening mechanisms,
scaled so that the peak flux is the same as the original model.  The parameters
for each line are given in the legend.  The obvious differences between the
solid and dashed lines in each panel (e.g., peak-to-center flux ratios and
velocity separation between peaks) illustrate how the effects of rotational
and thermal broadening differ from those of turbulent broadening in the context
of this toy model.
\label{fig:toy_line}}
\end{figure*}

{\it Temperature} --- The left panel of Figure~\ref{fig:toy_line} shows model
line profiles that illustrate the relationship between broadening due to
temperature and broadening due to a global turbulent linewidth.  The solid
line is the model line profile calculated using values from the best-fit
similarity solution: $T_{100}$=60\,K, $q$=0.5, and $\xi$=300\,m\,s$^{-1}$.
The dotted line represents the profile for a model with the same parameters
but no turbulent linewidth.  To create the dashed line profile, the temperature
was varied until the peak line flux of the model without turbulence matched the
peak flux of the original model with turbulence.  This sequence of line profiles
illustrates how thermal and turbulent broadening produce distinct effects on
the observable properties of the data, rather than being fully interchangeable.
Since increases in temperature increase the line flux throughout the disk while
turbulence simply redistributes the flux in frequency space, turbulence tends to
shift flux from the line peaks to the center, changing the shape of the
line.  As a result, the peak-to-center line flux ratios will be different for
line profiles with comparable widths and peak fluxes, depending on the relative
contributions of turbulence and temperature to line broadening.  The difference
between the best-fit 300\,m\,s$^{-1}$ turbulent linewidth in HD 163296 and a
comparable model without turbulence corresponds to a 30\% change in
peak-to-center line flux in the context of this toy model, which should be
easily distinguishable given the quality of our data.

As mentioned above, the temperature power law index $q$ is also found to be
degenerate with the turbulent linewidth.  The effects of this parameter are 
similar to those of the temperature normalization, illustrated in the left
panel of Figure \ref{fig:toy_line}: varying $q$ also alters the peak-to-center 
ratio of the line.  However, since the emission from both the peaks and the 
center is dominated by spatial scales of order 100 AU, the relevant spatial 
dynamic range is not large enough for variations of $q$ (within the 
uncertainties) to account for broadening on the scale of the derived turbulent 
linewidth.

{\it Inclination} --- The right panel of Figure~\ref{fig:toy_line} shows
model line profiles that illustrate the relationship between turbulent
broadening and rotational broadening due to inclination.  As in the left
panel, we first generate a model with no turbulent broadening (dotted line),
and then adjust the inclination until the peak flux matches the original line
model (dashed line).  In this case, the absence of turbulence once again
alters the peak-to-center line ratio.  But another obvious difference between
the turbulent model and the more inclined model without turbulence is that
rotational broadening through inclination alters the separation between the
line peaks in velocity space.  The sequence of toy models shows that the
contribution of the 300\,m\,s$^{-1}$ turbulent linewidth in the HD~163296
model corresponds to a 6$^\circ$ change in inclination, in addition to the
large depression of flux at the line center.  Since spatially resolved spectra
of the Keplerian rotation patterns in molecular lines allow for very precise
determination of circumstellar disk inclination \citep[to within a degree or
less, given the stellar mass and assuming axisymmetry; see discussion 
in][]{qi04}, turbulent line broadening at the level detected in HD~163296 
should be distinguishable from rotational broadening.  It should be noted that
rotational broadening includes the combined effects of inclination and stellar
mass since the two variables are almost precisely degenerate to within 
$\sim$10$^\circ$ (or tens of percent in mass): we cannot disentangle their 
effects.  Nevertheless, since we allow inclination to vary for the assumed 
stellar mass, and have demonstrated that inclination can be distinguished 
from turbulent broadening in the toy model line profiles, the uncertainty in 
stellar mass will not strongly affect our determination of the turbulent 
linewidth.

We have focused this discussion on HD~163296 to investigate the robustness of
the turbulent linewidth measurement rather than the upper limit for TW Hya.  
The situation is somewhat different for TW~Hya, since the face-on disk 
orientation does not result in splitting of the line profile, so diagnostics 
like peak-to-trough ratios and peak separations do not apply.  Nevertheless, 
the Keplerian rotation pattern -- the combination of inclination and stellar 
mass -- is still very precisely determined from the channel maps \citep{qi04}, 
and the temperature is well constrained by the brightness of the optically 
thick line.  As a result, we achieve a strong constraint on the turbulent 
linewidth for this system.  It should be noted that a turbulent linewidth 
comparable to that derived for HD~163296 would have a dramatic effect on the
line profile for this system, since a 300\,m/s turbulent linewidth would
represent roughly a 50\% perturbation on the observed 700\,m/s FWHM of the
CO(3-2) line.

\section{Discussion}
\label{sec:hires_discussion}

\subsection{Comparison with Theory}

The problem of accurately modeling and predicting the magnitude of velocity
fluctuations arising from magnetohydrodynamic (MHD) turbulence has proven
somewhat intractable, but a few general features seem to be agreed upon.  The 
difficulty hinges largely on derived values of the parameter $\alpha$ and how 
they relate to the expected magnitude of the turbulent linewidth.  The 
dimensionless parameter $\alpha$ was defined by \citet{sha73} to describe the 
effective viscosity in terms of a proportionality constant multiplied by the 
largest velocity and length scales on which turbulence may act: the scale 
height and the sound speed.  In mathematical terms, $\nu_\mathrm{eff} = \alpha 
c_s H$, where $\nu_\mathrm{eff}$ is the effective viscosity, $c_s$ is the 
sound speed, $H$ is the local scale height, and $\alpha$ is then an efficiency 
factor with a value $\leq$1.   The magnitude of turbulent velocity 
fluctuations depends both on the value of $\alpha$ and on how this efficiency 
factor is apportioned between the sound speed and the scale height.  If, for 
example, the majority of the power in turbulent fluctuations occurs at the 
length of the scale height, the velocity fluctuations can be as small as 
$\alpha c_s$ (perhaps augmented by a geometric factor of a few).  On the other 
hand, if the proportionality factor $\alpha$ applies evenly to the length and 
velocity scales in the problem, the turbulent fluctuations could be as large 
as $\sqrt{\alpha} c_s$ (again possibly modified by a geometric factor).  Since
there is no evidence of shocks that would point to sonic or supersonic
turbulence in circumstellar disks, it is unlikely that the turbulent velocity
fluctuations would be much larger than $\sqrt{\alpha} c_s$.

It is clear that shearing box simulations of MHD turbulence with zero net
magnetic field flux do not give reliable values of the viscosity parameter
$\alpha$ due to numerical dissipation, which results in values of $\alpha$ that
depend on resolution \citep{pes07,fro07}.  This is mitigated by the use of 
more realistic simulation conditions including vertical stratification 
\citep[e.g.,][]{sto96,fle03,dav10,fla10}, but the resulting dependence of 
$\alpha$ and therefore turbulent velocity fluctuations on the magnitude of the 
magnetic field is troublesome, since magnetic field strengths and geometries 
in circumstellar disks are unconstrained by observations 
\citep[e.g.,][]{hug09b}.  The inclusion of vertical gravity has also been 
shown to lead to convergence \citep[e.g.][]{dav10,fla10}.  It is perhaps 
instructive to investigate how $\alpha$ may be related to the observed 
turbulent linewidth in the context of shearing-box simulations.  Quantities 
typically reported for shearing boxes include the shear stress $-B_x B_y / 4 
\pi$, where $B_x$ and $B_y$ are the magnetic field along the $x$ and $y$ 
directions, respectively; Reynolds stress $\rho v_x \delta v_y$, where $\rho$ 
is density, $v_x$ is the $x$ component of velocity, and $\delta v_y$ is the 
$y$ component of velocity without shear; and the magnitude of velocity 
fluctuations along the different dimensions of the simulation $\rho 
v_{x,y,z}^2/2$. Each of these quantities is routinely normalized by the 
initial midplane pressure, $P_0$ \citep[see, e.g.,][]{sto96,dav10}.  The 
total stress $T$ is then the sum of the Maxwell and Reynolds stresses, and 
is directly proportional to the viscosity parameter $\alpha$, according to the 
relationship $T = \alpha \rho c_s^2$, where $c_s$ is the sound speed.  It is 
then straightforward to relate $\alpha$ to the turbulent velocity as a 
function of sound speed:
\begin{eqnarray}
\alpha = \frac{T}{\rho v_\mathrm{turb}^2} \frac{v_\mathrm{turb}^2}{c_s^2} \\
\alpha = \frac{T/P_0}{2(\rho v_z^2 \cos^2 i+\rho v_{x,y}^2 \sin^2 i)/2P_0} 
\frac{v_\mathrm{turb}^2}{c_s^2}
\end{eqnarray}
where $v_\mathrm{turb}$ is the measured turbulent linewidth (typically the 
FWHM calculated from $\xi$, which is the $1/e$ half-width) and $i$ is the
disk inclination.  Using reported values of the relevant simulation 
parameters from \citet{dav10}, $\alpha \sim 3 (v_\mathrm{turb}/c_s)^2$ for a 
face-on disk, and $\alpha \sim 1.5 (v_\mathrm{turb}/c_s)^2$ for an edge-on
disk.  For the value of $\alpha \sim 0.01$ derived in \citet{dav10}, this
predicts a midplane turbulent linewidth of roughly 6-8\% of the sound speed,
depending on the source geometry.

Observational attempts at constraining the value of $\alpha$ in circumstellar
disks, while generally quite uncertain, seem to cluster near values
of 10$^{-2}$, with large scatter \citep[e.g.,][]{har98,and09a}.  If the
velocity fluctuations are estimated as $\sqrt{\alpha} c_s$, this result would
imply velocity fluctuations up to 10\% of the sound speed near the midplane,
although this global value is likely to vary widely depending on local
conditions that affect the ionization fraction and the coupling of ions and
neutrals.  A theoretical comparison between circumstellar disks and the
Taylor-Couette flow by \citet{her05} finds that the 100\,m\,s$^{-1}$ turbulent
linewidth (of order $\lesssim$30\% of the sound speed) derived from low
spectral resolution observations of DM Tau by \citet{gui98} are on par with
expectations from laboratory measurements by \citet{dub04}.  There is some
evidence, both from the study of FU Ori objects \citep{har04} and global
MHD simulations of stratified disks \citep[e.g.][]{fro06}, that the turbulent
linewidth may be larger at several scale heights above the midplane of the
disk, perhaps up to 40\% of the sound speed.  While these estimates affect the
local value of $\alpha$, it should also be noted that there are global disk
features that may also affect the measured turbulent linewidth.  In the dense
inner disk, dead zones may form near the midplane where the ionization fraction
is too low to support the MRI \citep[e.g.][]{san00}.  However, these should not 
be relevant for our observations since the predicted extent of a dead zone in 
these disks is interior to a radius of 5-25\,AU for TW Hya and 10-63\,AU for 
HD 163296, depending on the assumed grain properties (N. Turner 2010, private 
communication); these size scales are significantly smaller than the linear 
scale probed by our observations.

In this context, HD~163296 seems to fit in fairly seamlessly, with a turbulent
linewidth of 300\,m\,s$^{-1}$ corresponding to about 40\% of the sound speed
at the size scales probed by our data (if 1\farcs5 corresponds to $\sim$180\,AU
at a distance of 122\,pc).  If the turbulent linewidth drops by a factor of
a few between the upper layers probed by the CO(3-2) line and the midplane 
\citep[as predicted by, e.g.,][]{fro06}, this implies a turbulent linewidth of 
$\sim$0.1\,$c_s$ near the midplane, consistent with $\alpha \sim 0.01$.  The 
lower turbulent linewidth in TW~Hya, $\lesssim$10\% of the local sound speed 
at the scale of our observations (80\,AU at 51\,pc), could be associated with a 
lower global value of $\alpha$, although the reason for such a difference is 
unclear.  Following \citet{and09a}, the best-fit similarity solution model 
parameters imply $\alpha$ = 0.03 for TW~Hya and 0.009 for HD~163296, although 
these numbers are quite uncertain given the reliance on the optically thick 
CO(3-2) tracer for the determination of density.  The D'Alessio et al.~models 
use $\alpha$ = 0.0018 for TW~Hya and 0.003 for HD~163296.  Perhaps the only 
conclusion that should be drawn from this estimate is that our observations 
are consistent with estimates that place $\alpha$ in the range of 
$10^{-2}$-10$^{-3}$.  With a sample of only two sources and with a wide range 
of theoretical results and approaches to the study of turbulence, it can be 
difficult to compare our results to theoretical predictions in a detailed way.  
Nevertheless, the detection of a turbulent linewidth in the HD~163296 system 
and the upper limit on turbulence in the TW~Hya system are suggestive.  There 
are several potential explanations for this difference rooted in theoretical 
studies of protoplanetary disk turbulence.

{\it Inclination and Vertical Structure} --- In general, the CO(3-2) line
emission from these systems is expected to be optically thick, and therefore
will arise from the tenuous upper disk layers several scale heights above
the midplane.  However, this simple picture can be complicated by geometry
and velocity: the combination of inclination and rotational line broadening in
HD~163296 will permit the escape of radiation from deeper layers of the disk.
As a result, a different vertical height may be probed in the HD~163296 system
than in TW~Hya.  Naively, it would be expected that the turbulent linewidth in
TW~Hya should be {\it larger} as a result, since turbulence is predicted to
become stronger farther from the midplane.  It is difficult to calculate 
accurately the height above the midplane from which the CO(3-2) emission
originates, both because it varies by position across the disk and velocity
offset across the line, and because it depends on very poorly constrained 
quantities like the vertical gradients in gas temperature and density.  
Nevertheless, a rough estimate at line center for the D'Alessio et al.~models
places the origin of the emission at approximately 2 scale heights above the 
midplane for TW Hya and 3 for HD~163296 at the radii of 80 and 180 AU 
corresponding to the respective linear resolution of the observations for 
each source.  This suggests that the difference in turbulent linewidth for 
the two sources could be due to the different physical regions probed by the
line.  But given the uncertainties, the height of formation is not constrained
stringently enough to definitively demonstrate that this is the case.  It is 
also possible that poor coupling of ions to neutrals in the low-density 
uppermost disk layers could inhibit the detection of turbulence even if it 
is present.  This may also depend on the relative amounts of small dust grains 
in the upper layers of the two disks, since small grains are more adept at 
absorbing free electrons.  

{\it Stellar Mass and Ionizing Flux} --- One of the factors determining the
extent of turbulent regions in circumstellar disks is the magnitude of
ionizing X-ray flux from the star \citep[e.g.,][]{gam96,ige99}.  With stellar
masses differing by a factor of 4, the radiation (and therefore ionization)
properties of TW~Hya and HD~163296 are likely to be quite different.  Despite
the tendency for X-ray flux to be lower for intermediate-mass than low-mass
stars of comparable ages, HD~163296 has the largest X-ray flux of the sample
of 13 Herbig Ae stars studied by \citet{hub09}, comparable to that of
lower-mass T Tauri stars.  Nevertheless, its measured X-ray luminosity of
$4.0 \times 10^{29}$\,erg\,s$^{-1}$ is still lower than that of TW~Hya, which
is estimated as $2.0 \times 10^{30}$\,erg\,s$^{-1}$ by \citet{kas99}.  Given
the comparable disk densities, by X-ray luminosity alone, TW~Hya should be the
more active disk.  However, since we are observing these disks on relatively
large spatial scales ($\sim$80\,AU for TW~Hya and $\sim$180\,AU for HD~163296,
with 1\farcs5 resolution viewed from their respective 51 and 122\,pc
distances), the ionization at these distances may instead be dominated by
cosmic rays.  It is extremely difficult to determine how the cosmic ray
environment of these two sources might compare; if HD~163296 were located in
a region of greater cosmic ray activity, that could account for the greater
turbulent linewidth observed for this system.

{\it Evolutionary State} --- One of the most obvious differences between
TW~Hya and HD~163296 is their respective evolutionary states.  TW~Hya is a
10\,Myr-old transition disk with an inner cavity of $\sim$4\,AU radius, while
HD~163296 is a primordial disk with an inner radius consistent with the
expected $\sim$0.4\,AU extent of the dust destruction zone \citep[see,
e.g.,][]{ise07}.  It is possible that X-ray ionization and the MRI might
operate differently at different stages of evolution; one example is the
inside-out MRI clearing proposed by \citet{chi07} to explain the cavities in
transition disks.  In their scenario, the mass accretion rate onto the star
can be explained entirely by the MRI operating on the disk inner rim, and
requires no resupply from the outer disk; they therefore require little to no
turbulent viscosity in the outer disk to explain the observed accretion rates
in transitional systems.  However, they note that their theory cannot be
readily applied to primordial systems, leaving the viscous transport mechanism
responsible for large accretion rates at earlier stages unexplained.  There
is no reason to expect MRI turbulence in the outer disk to ``shut off'' when
a gap is opened, so while our observation of a small turbulent linewidth in
the TW~Hya system is consistent with the \citet{chi07} hypothesis, it is still
surprising that the turbulent linewidth in HD~163296 should be so much larger.
Another possibility unrelated to the MRI is that HD~163296 is still 
experiencing infall onto the disk from a remnant envelope.  Since the high
optical depth of the CO(3-2) line implies that most of the emission arises 
from the upper layers of the disk, such a scenario could mimic the signature of
a turbulent linewidth that we observe; however, there is no observational 
evidence for an envelope in this system.  We plan to address this possibility 
more thoroughly in a follow-up paper modeling multiple lines with lower 
optical depths that probe deeper towards the disk midplane.

\subsection{Implications for Planet Formation}

The presence of subsonic turbulence in protoplanetary accretion disks --
likely substantially subsonic in the midplane -- is consistent with the
observations presented in this study.  Subsonic turbulence has important
implications for the formation and evolution of young planetary systems.
One series of papers \citep{pap03,pap04,nel03,nel04} explores in detail the
effects of turbulence on planet-forming disks.  Their cylindrical models of
turbulent disks have an average $\alpha$ in the range of $10^{-2}$-$10^{-3}$,
but they demonstrate that the realistic implementation of turbulence results
in different effects than are seen in laminar disk simulations with comparable
values of $\alpha$ incorporated as an anomalous Navier-Stokes viscosity.  They
show that for massive planets, turbulence can widen and deepen the gap opened
by massive protoplanets, and may reduce the accretion rate onto the
protoplanet.  For the case of migrating low-mass planet cores, the presence
of turbulence in the disk can slow or even reverse the migration rate,
converting the monotonic inward motion of the planet into a random walk.
The presence of dead zones in the radial direction may also act to halt
migration and encourage the survival and growth of protoplanets
\citep[e.g.][]{mat09}.  Another important proposed effect of subsonic
turbulence is to aid in concentrating planetesimals to allow them to collapse
gravitationally \citep{joh07}.  MHD turbulence on these scales can also reduce
the strength of the gravitational instability and reduce disk fragmentation
\citep{fro05}.  There is also substantial literature on the effects of
turbulence on dust settling and grain growth \citep[e.g.][]{joh05,car06,
cie07,bal09,fro09}.

Although it is difficult to compare the properties of the simulations directly
with our observations, the generic features of these models
($\alpha$=$10^{-2}$-$10^{-3}$, subsonic turbulence even in the upper disk
layers) are globally consistent with the derived properties of turbulence in
the disks around HD~163296 and TW~Hya, indicating that these effects are
likely to play a role in planet formation.

\subsection{Future Directions}

The most obvious improvement to our method would be to include additional
spectral lines from different transitions or isotopologues of the CO molecule
in order to provide independent constraints on the gas temperature.  While
this would necessarily introduce additional parameters into the model (i.e.,
to describe the vertical distribution of temperature and turbulence,
as well as a consistent density distribution to properly account for the
line opacity), the addition of several lines that are resolved in the spectral
and spatial domains would more firmly constrain the models.  It might also
provide direct measurements of the vertical profile of the turbulent velocity
structure.  \citet{dar03} and \citet{pan08} provide examples of studies that
use multiple molecular lines to study the vertical structure of density and
temperature in circumstellar disks; these techniques could be extended to
constrain the turbulent linewidths in similar systems.

Another possibility is to observe ions rather than neutral species.
This would eliminate complications introduced by the interaction between
ions and neutrals, and would more directly probe the turbulent motions of
the charged gas.

Even with the current set of observations, greater sensitivity would be
extremely valuable in constraining the turbulent linewidth, since the
distinctions between turbulent and thermal broadening are subtle (see
Section~\ref{sec:degen}).  The vast improvements in sensitivity provided by
the Atacama Large Millimeter Array will permit significantly better modeling
of the velocity structure of young disks.  Such data will also allow us to
firmly rule out deviations from perfect Keplerian rotation that could 
complicate the derivation of turbulent linewidth.  In addition, higher 
sensitivity combined with a greater spatial dynamic range will allow for the 
investigation of radial variations in the turbulent linewidth.

\section{Summary and Conclusions}
\label{sec:hires_summary}

We have obtained the first spatially resolved observations of molecular line
emission from two nearby circumstellar disks with spectral resolution finer
than the expected turbulent linewidth.  We fit these high spectral resolution
observations of the CO(3-2) line emission using two well-tested models of
circumstellar disk structure, and derive a turbulent linewidth of
$\sim$300\,m\,s$^{-1}$ for the disk around HD~163296 and
$\lesssim$40\,m\,s$^{-1}$ for the disk around TW~Hya.  The results are
consistent for the two model classes despite their different treatments of
the temperature structure of the disk, which is significant since thermal
broadening is closely related to turbulent broadening.  The magnitude of
turbulent velocity fluctuations implied by these results -- up to tens of
percent of the sound speed -- is broadly consistent with theoretical
predictions for MRI turbulence in the upper layers of circumstellar disks,
although it is unclear why the linewidth should be lower for TW~Hya than for
HD~163296.

These results demonstrate the potential of this method for
constraining theories of the magnitude and spatial distribution of turbulence.
Future observations with greater sensitivity, perhaps incorporating different
molecular line species and isotopologues of the same molecule, have the
potential to vastly improve our ability to characterize turbulence in these
systems.

\acknowledgments
The authors thank Shin-Yi Lin for collaboration on collection of the 
extended TW Hya observations.  We also thank Ramesh Narayan for helpful 
conversations about comparing the data with theoretical simulations, as well 
as Shane Davis for providing several quantities that were not tabulated in 
his paper. Partial support for this work was provided by NASA Origins of 
Solar Systems Program Grant NAG5-11777.  A.~M.~H. acknowledges support from 
a National Science Foundation Graduate Research Fellowship.

\appendix

\section{Channel Maps}
\label{sec:hires_appendix}

\begin{figure}[t]
\epsscale{1.0}
\plotone{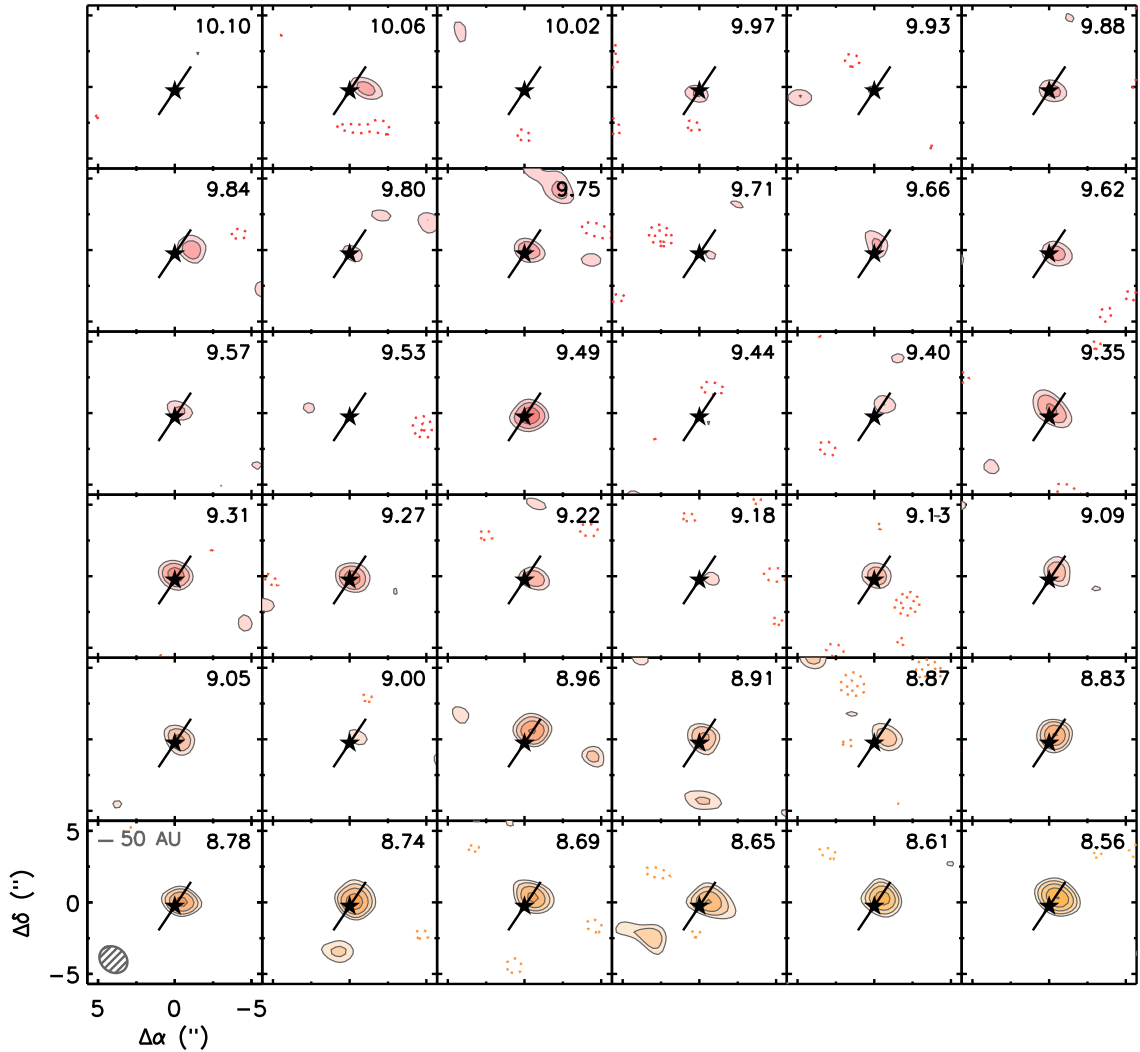}
\figcaption{
Channel maps of the CO(3-2) line emission from the disk around HD~163296.  
The LSR velocity is indicated in the upper right of each channel, while the 
synthesized beam size and orientation (2\farcs0$\times$1\farcs7 at a position 
angle of 41$^\circ$) is indicated in the lower left panel.  The contour
levels start at 2$\sigma$ and increase by factors of $\sqrt{2}$, where $\sigma$
is the rms noise of 0.6\,Jy\,beam$^{-1}$.  The star symbol indicates the disk
center while the dark solid line indicates the disk position angle as
determined by CO fitting in Section~\ref{sec:hires_analysis}.
\label{fig:hd163296_chmap}}
\end{figure}

\begin{figure}[t]
\epsscale{0.8}
\plotone{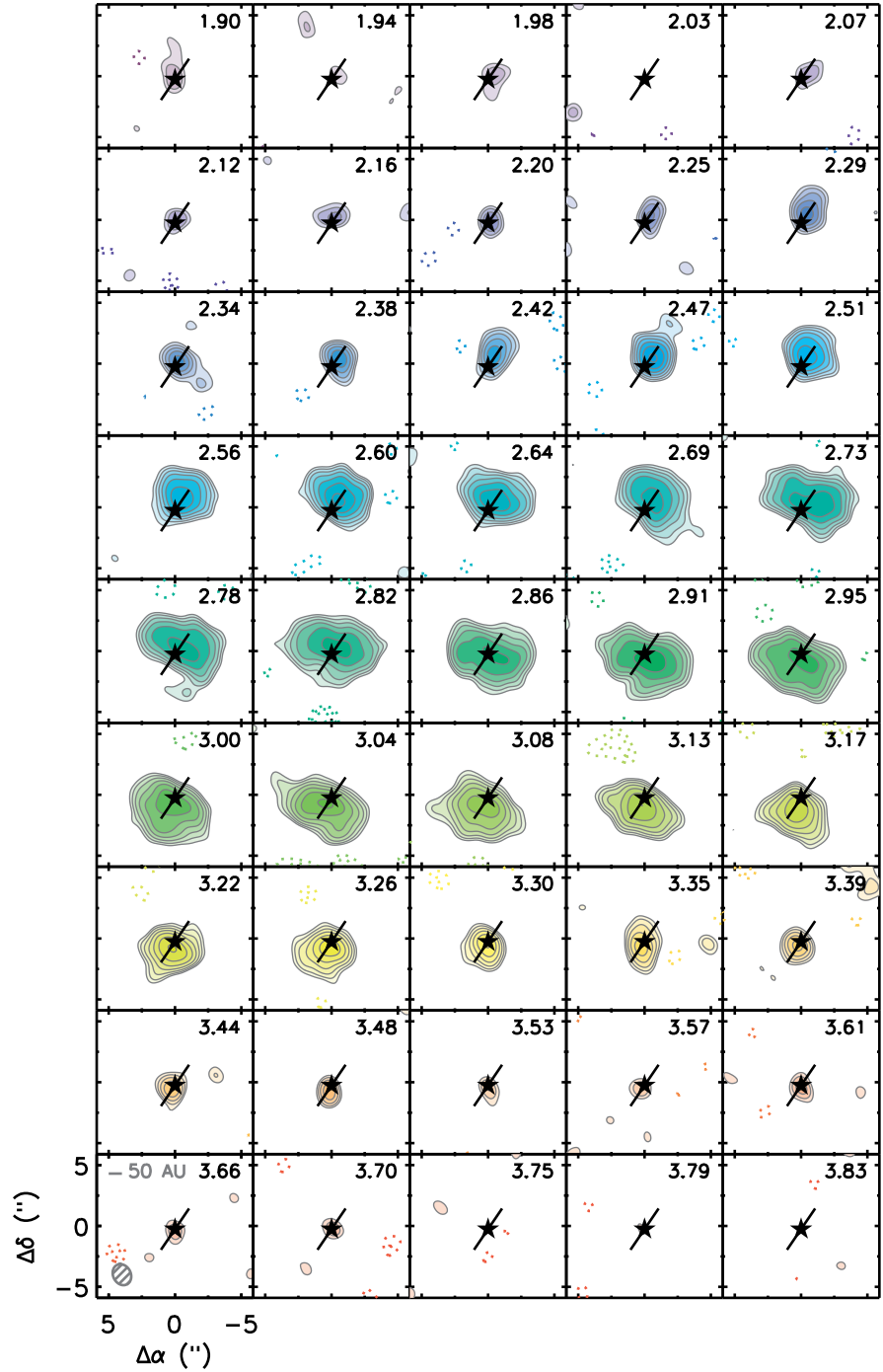}
\figcaption{Same as Figure~\ref{fig:hd163296_chmap} above, but for TW~Hya.
The beam size is 1\farcs9$\times$1\farcs4 at a position angle of 21$^\circ$ 
and the contour levels are the same as in Figure~\ref{fig:hd163296_chmap}.
\label{fig:twhya_chmap}}
\end{figure}

Figures~\ref{fig:hd163296_chmap} and \ref{fig:twhya_chmap} show the full 
channel maps for the high spectral resolution observations of the CO(3-2)
emission from the disks around TW Hya and HD 163296.  The line overlaid
on the emission indicates the disk major axis.  The TW Hya and HD~163296 maps 
have been imaged with Gaussian tapers of 1\farcs2 and 1\farcs0, respectively,
to bring out the large-scale emission.

\begin{figure}[t]
\epsscale{1.0}
\plotone{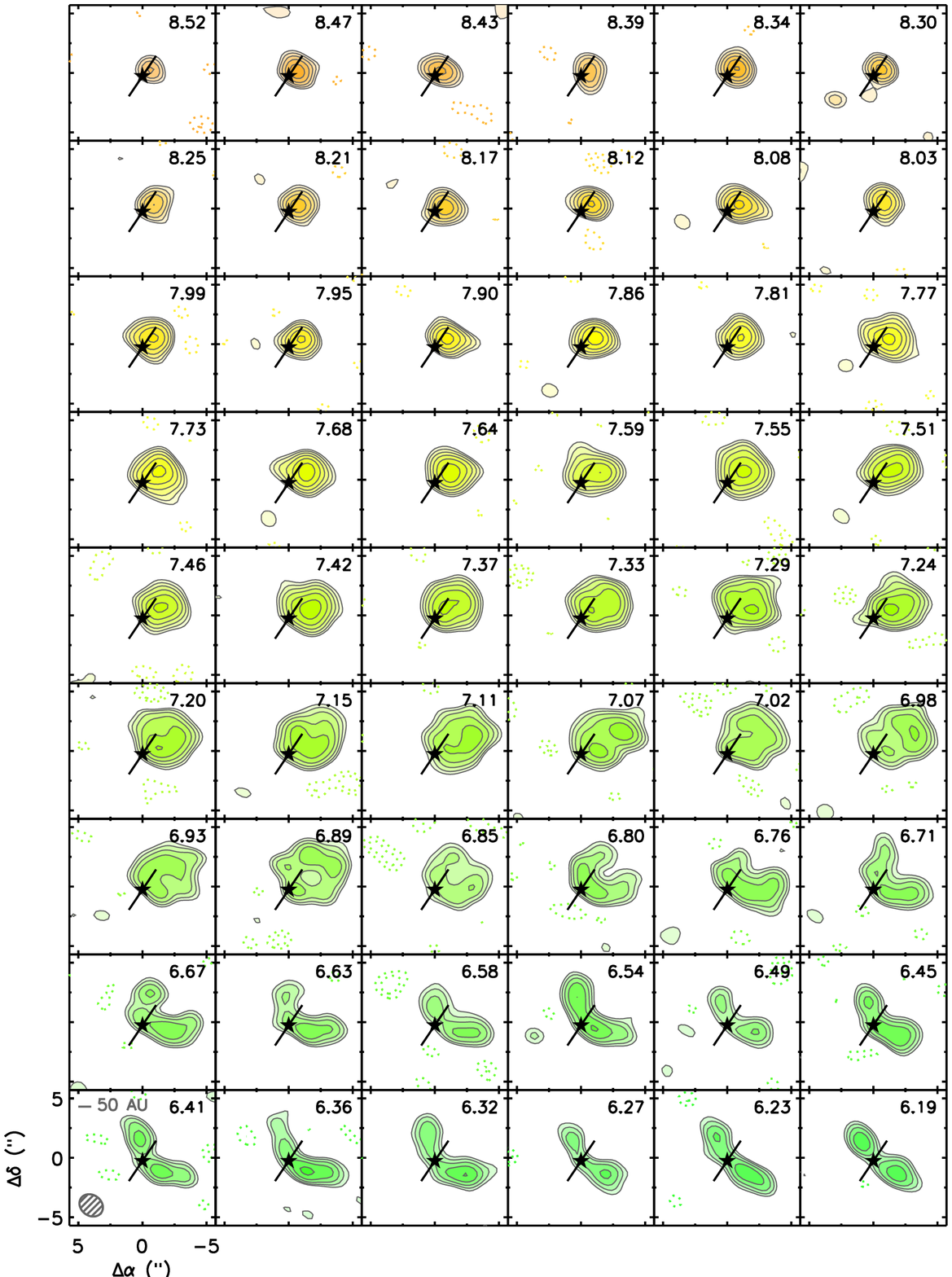}
\end{figure}

\begin{figure}[t]
\epsscale{1.0}
\plotone{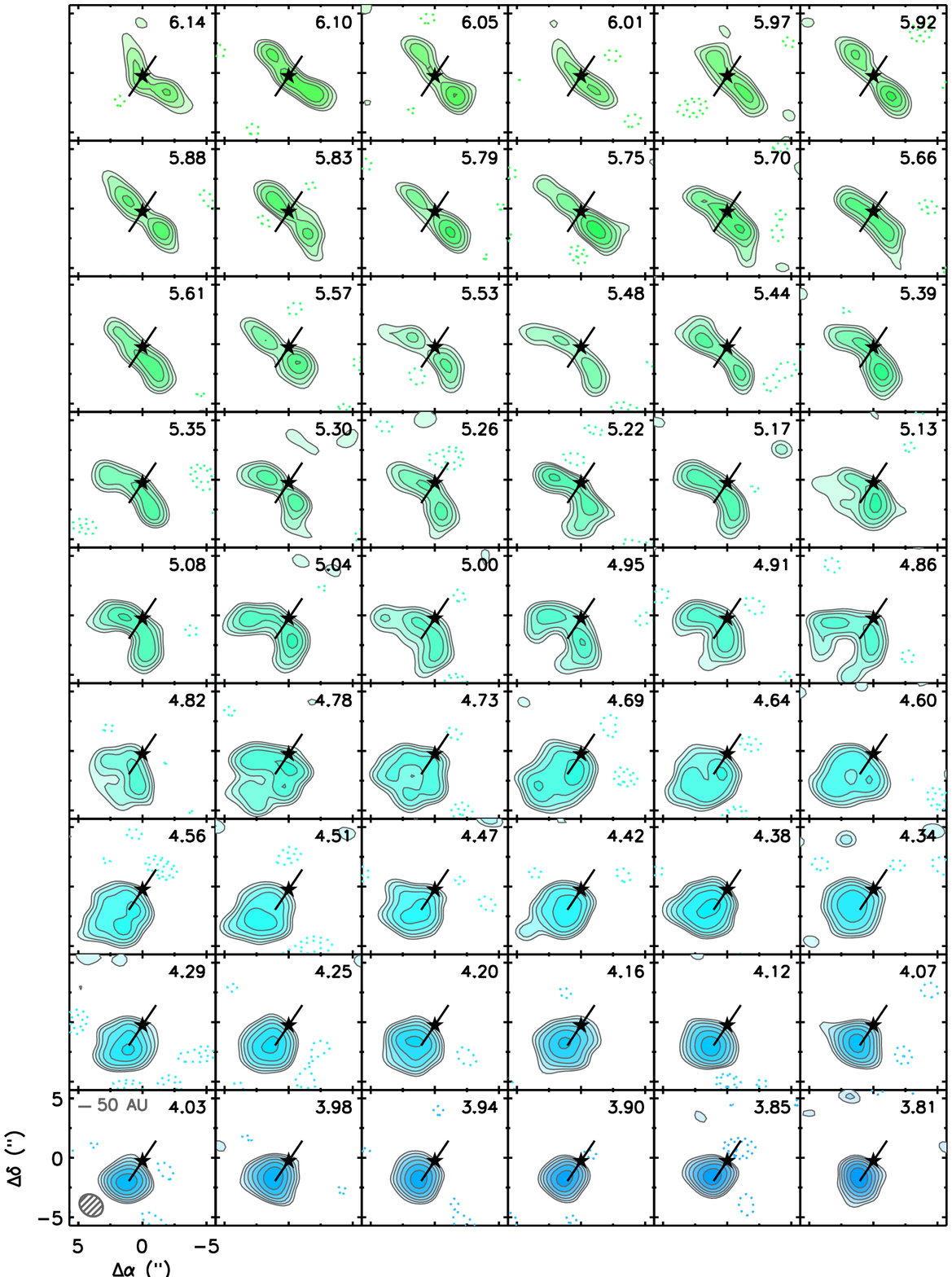}
\end{figure}

\begin{figure}[t]
\epsscale{1.0}
\plotone{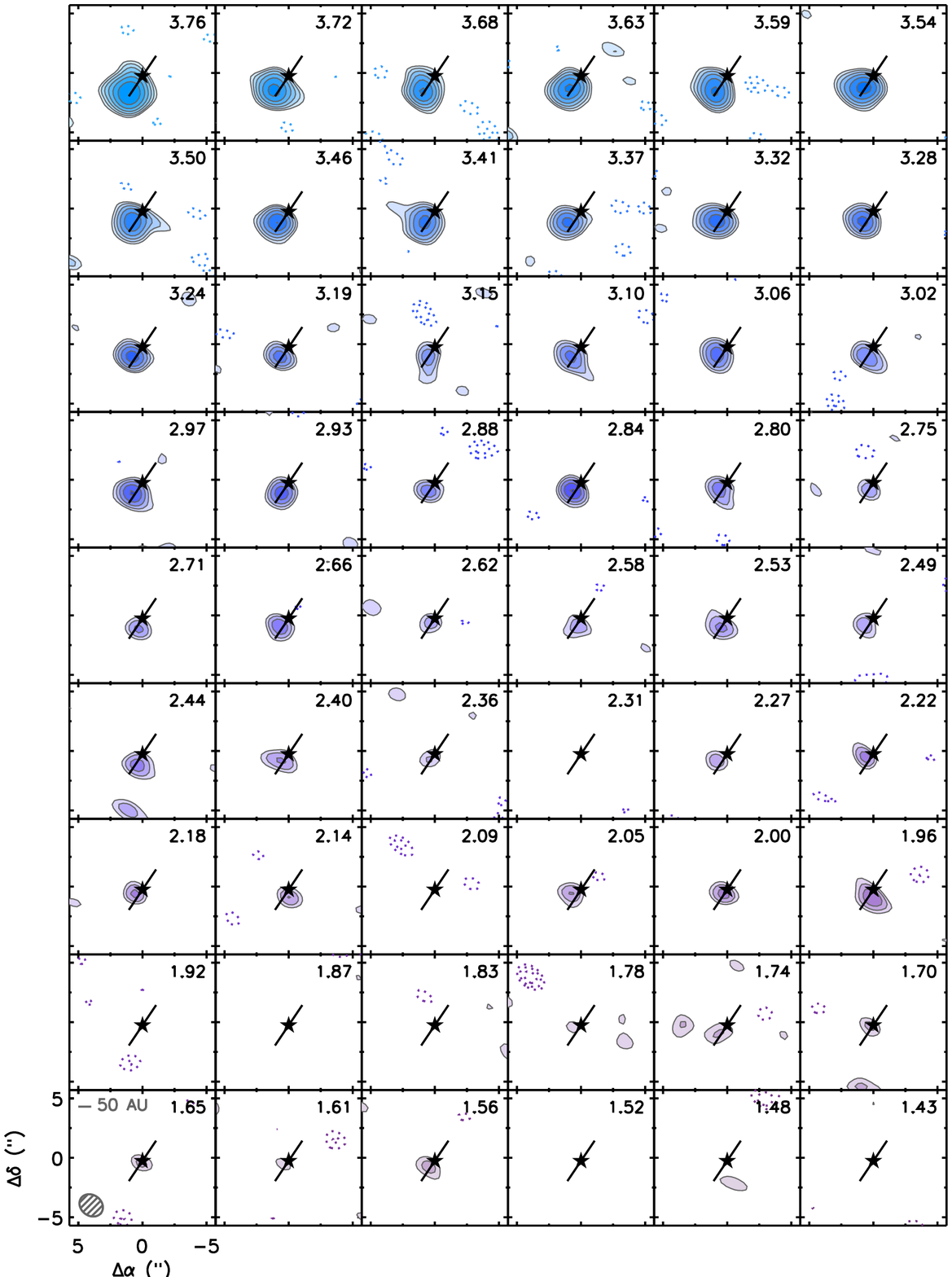}
\end{figure}

\bibliography{ms}

\begin{thebibliography}{72}
\expandafter\ifx\csname natexlab\endcsname\relax\def\natexlab#1{#1}\fi

\bibitem[{{Adams} \& {Shu}(1986)}]{ada86}
{Adams}, F.~C. \& {Shu}, F.~H. 1986, \apj, 308, 836

\bibitem[{{Andrews} \& {Williams}(2007)}]{and07}
{Andrews}, S.~M. \& {Williams}, J.~P. 2007, \apj, 659, 705

\bibitem[{{Andrews} {et~al.}(2009){Andrews}, {Wilner}, {Hughes}, {Qi}, \&
  {Dullemond}}]{and09a}
{Andrews}, S.~M., {Wilner}, D.~J., {Hughes}, A.~M., {Qi}, C., \& {Dullemond},
  C.~P. 2009, \apj, 700, 1502

\bibitem[{{Balbus} \& {Hawley}(1991)}]{bal91}
{Balbus}, S.~A. \& {Hawley}, J.~F. 1991, \apj, 376, 214

\bibitem[{{Balbus} \& {Hawley}(1998)}]{bal98}
---. 1998, Reviews of Modern Physics, 70, 1

\bibitem[{{Balsara} {et~al.}(2009){Balsara}, {Tilley}, {Rettig}, \&
  {Brittain}}]{bal09}
{Balsara}, D.~S., {Tilley}, D.~A., {Rettig}, T., \& {Brittain}, S.~D. 2009,
  \mnras, 397, 24

\bibitem[{{Beckwith} \& {Sargent}(1991)}]{bec91}
{Beckwith}, S.~V.~W. \& {Sargent}, A.~I. 1991, \apj, 381, 250

\bibitem[{{Beckwith} {et~al.}(1990){Beckwith}, {Sargent}, {Chini}, \&
  {Guesten}}]{bec90}
{Beckwith}, S.~V.~W., {Sargent}, A.~I., {Chini}, R.~S., \& {Guesten}, R. 1990,
  \aj, 99, 924

\bibitem[{{Boss}(2004)}]{bos04}
{Boss}, A.~P. 2004, \apj, 616, 1265

\bibitem[{{Calvet} {et~al.}(2002){Calvet}, {D'Alessio}, {Hartmann}, {Wilner},
  {Walsh}, \& {Sitko}}]{cal02}
{Calvet}, N., {D'Alessio}, P., {Hartmann}, L., {Wilner}, D., {Walsh}, A., \&
  {Sitko}, M. 2002, \apj, 568, 1008

\bibitem[{{Calvet} {et~al.}(2005){Calvet}, {D'Alessio}, {Watson},
  {Franco-Hern{\'a}ndez}, {Furlan}, {Green}, {Sutter}, {Forrest}, {Hartmann},
  {Uchida}, {Keller}, {Sargent}, {Najita}, {Herter}, {Barry}, \&
  {Hall}}]{cal05}
{Calvet}, N., {D'Alessio}, P., {Watson}, D.~M., {Franco-Hern{\'a}ndez}, R.,
  {Furlan}, E., {Green}, J., {Sutter}, P.~M., {Forrest}, W.~J., {Hartmann}, L.,
  {Uchida}, K.~I., {Keller}, L.~D., {Sargent}, B., {Najita}, J., {Herter},
  T.~L., {Barry}, D.~J., \& {Hall}, P. 2005, \apjl, 630, L185

\bibitem[{{Carballido} {et~al.}(2006){Carballido}, {Fromang}, \&
  {Papaloizou}}]{car06}
{Carballido}, A., {Fromang}, S., \& {Papaloizou}, J. 2006, \mnras, 373, 1633

\bibitem[{{Carr} {et~al.}(2004){Carr}, {Tokunaga}, \& {Najita}}]{car04}
{Carr}, J.~S., {Tokunaga}, A.~T., \& {Najita}, J. 2004, \apj, 603, 213

\bibitem[{{Chiang} \& {Murray-Clay}(2007)}]{chi07}
{Chiang}, E. \& {Murray-Clay}, R. 2007, Nature Physics, 3, 604

\bibitem[{{Ciesla}(2007)}]{cie07}
{Ciesla}, F.~J. 2007, \apjl, 654, L159

\bibitem[{{D'Alessio} {et~al.}(2001){D'Alessio}, {Calvet}, \&
  {Hartmann}}]{dal01}
{D'Alessio}, P., {Calvet}, N., \& {Hartmann}, L. 2001, \apj, 553, 321

\bibitem[{{D'Alessio} {et~al.}(2006){D'Alessio}, {Calvet}, {Hartmann},
  {Franco-Hern{\'a}ndez}, \& {Serv{\'{\i}}n}}]{dal06}
{D'Alessio}, P., {Calvet}, N., {Hartmann}, L., {Franco-Hern{\'a}ndez}, R., \&
  {Serv{\'{\i}}n}, H. 2006, \apj, 638, 314

\bibitem[{{D'Alessio} {et~al.}(1999){D'Alessio}, {Calvet}, {Hartmann},
  {Lizano}, \& {Cant{\'o}}}]{dal99}
{D'Alessio}, P., {Calvet}, N., {Hartmann}, L., {Lizano}, S., \& {Cant{\'o}}, J.
  1999, \apj, 527, 893

\bibitem[{{D'Alessio} {et~al.}(1998){D'Alessio}, {Canto}, {Calvet}, \&
  {Lizano}}]{dal98}
{D'Alessio}, P., {Canto}, J., {Calvet}, N., \& {Lizano}, S. 1998, \apj, 500,
  411

\bibitem[{{Dartois} {et~al.}(2003){Dartois}, {Dutrey}, \& {Guilloteau}}]{dar03}
{Dartois}, E., {Dutrey}, A., \& {Guilloteau}, S. 2003, \aap, 399, 773

\bibitem[{{Davis} {et~al.}(2010){Davis}, {Stone}, \& {Pessah}}]{dav10}
{Davis}, S.~W., {Stone}, J.~M., \& {Pessah}, M.~E. 2010, \apj, 713, 52

\bibitem[{{Dent} {et~al.}(2005){Dent}, {Greaves}, \& {Coulson}}]{den05}
{Dent}, W.~R.~F., {Greaves}, J.~S., \& {Coulson}, I.~M. 2005, \mnras, 359, 663

\bibitem[{{Dubrulle} {et~al.}(2005){Dubrulle}, {Dauchot}, {Daviaud},
  {Longaretti}, {Richard}, \& {Zahn}}]{dub04}
{Dubrulle}, B., {Dauchot}, O., {Daviaud}, F., {Longaretti}, P., {Richard}, D.,
  \& {Zahn}, J. 2005, Physics of Fluids, 17, 095103

\bibitem[{{Dutrey} {et~al.}(1994){Dutrey}, {Guilloteau}, \& {Simon}}]{dut94}
{Dutrey}, A., {Guilloteau}, S., \& {Simon}, M. 1994, \aap, 286, 149

\bibitem[{{Flaig} {et~al.}(2010){Flaig}, {Kley}, \& {Kissmann}}]{fla10}
{Flaig}, M., {Kley}, W., \& {Kissmann}, R. 2010, \mnras, 1450

\bibitem[{{Fleming} \& {Stone}(2003)}]{fle03}
{Fleming}, T. \& {Stone}, J.~M. 2003, \apj, 585, 908

\bibitem[{{Fromang}(2005)}]{fro05}
{Fromang}, S. 2005, \aap, 441, 1

\bibitem[{{Fromang} \& {Nelson}(2006)}]{fro06}
{Fromang}, S. \& {Nelson}, R.~P. 2006, \aap, 457, 343

\bibitem[{{Fromang} \& {Nelson}(2009)}]{fro09}
---. 2009, \aap, 496, 597

\bibitem[{{Fromang} \& {Papaloizou}(2007)}]{fro07}
{Fromang}, S. \& {Papaloizou}, J. 2007, \aap, 476, 1113

\bibitem[{{Gammie}(1996)}]{gam96}
{Gammie}, C.~F. 1996, \apj, 457, 355

\bibitem[{{Gammie} \& {Johnson}(2005)}]{gam05}
{Gammie}, C.~F. \& {Johnson}, B.~M. 2005, in Astronomical Society of the
  Pacific Conference Series, Vol. 341, Chondrites and the Protoplanetary Disk,
  ed. A.~N. {Krot}, E.~R.~D. {Scott}, \& B.~{Reipurth}, 145--+

\bibitem[{{Grady} {et~al.}(2000){Grady}, {Devine}, {Woodgate}, {Kimble},
  {Bruhweiler}, {Boggess}, {Linsky}, {Plait}, {Clampin}, \& {Kalas}}]{gra00}
{Grady}, C.~A., {Devine}, D., {Woodgate}, B., {Kimble}, R., {Bruhweiler},
  F.~C., {Boggess}, A., {Linsky}, J.~L., {Plait}, P., {Clampin}, M., \&
  {Kalas}, P. 2000, \apj, 544, 895

\bibitem[{{Guilloteau} \& {Dutrey}(1998)}]{gui98}
{Guilloteau}, S. \& {Dutrey}, A. 1998, \aap, 339, 467

\bibitem[{{Hartmann} {et~al.}(1998){Hartmann}, {Calvet}, {Gullbring}, \&
  {D'Alessio}}]{har98}
{Hartmann}, L., {Calvet}, N., {Gullbring}, E., \& {D'Alessio}, P. 1998, \apj,
  495, 385

\bibitem[{{Hartmann} {et~al.}(2004){Hartmann}, {Hinkle}, \& {Calvet}}]{har04}
{Hartmann}, L., {Hinkle}, K., \& {Calvet}, N. 2004, \apj, 609, 906

\bibitem[{{Hersant} {et~al.}(2005){Hersant}, {Dubrulle}, \& {Hur{\'e}}}]{her05}
{Hersant}, F., {Dubrulle}, B., \& {Hur{\'e}}, J. 2005, \aap, 429, 531

\bibitem[{{Ho} \& {Townes}(1983)}]{ho83}
{Ho}, P.~T.~P. \& {Townes}, C.~H. 1983, \araa, 21, 239

\bibitem[{{Hogerheijde} \& {van der Tak}(2000)}]{hog00}
{Hogerheijde}, M.~R. \& {van der Tak}, F.~F.~S. 2000, \aap, 362, 697

\bibitem[{{Hubrig} {et~al.}(2009){Hubrig}, {Stelzer}, {Sch{\"o}ller}, {Grady},
  {Sch{\"u}tz}, {Pogodin}, {Cur{\'e}}, {Hamaguchi}, \& {Yudin}}]{hub09}
{Hubrig}, S., {Stelzer}, B., {Sch{\"o}ller}, M., {Grady}, C., {Sch{\"u}tz}, O.,
  {Pogodin}, M.~A., {Cur{\'e}}, M., {Hamaguchi}, K., \& {Yudin}, R.~V. 2009,
  \aap, 502, 283

\bibitem[{{Hughes} {et~al.}(2007){Hughes}, {Wilner}, {Calvet}, {D'Alessio},
  {Claussen}, \& {Hogerheijde}}]{hug07}
{Hughes}, A.~M., {Wilner}, D.~J., {Calvet}, N., {D'Alessio}, P., {Claussen},
  M.~J., \& {Hogerheijde}, M.~R. 2007, \apj, 664, 536

\bibitem[{{Hughes} {et~al.}(2009{\natexlab{a}}){Hughes}, {Wilner}, {Cho},
  {Marrone}, {Lazarian}, {Andrews}, \& {Rao}}]{hug09}
{Hughes}, A.~M., {Wilner}, D.~J., {Cho}, J., {Marrone}, D.~P., {Lazarian}, A.,
  {Andrews}, S.~M., \& {Rao}, R. 2009{\natexlab{a}}, \apj, 704, 1204

\bibitem[{{Hughes} {et~al.}(2009{\natexlab{b}}){Hughes}, {Wilner}, {Cho},
  {Marrone}, {Lazarian}, {Andrews}, \& {Rao}}]{hug09b}
---. 2009{\natexlab{b}}, \apj, 704, 1204

\bibitem[{{Hughes} {et~al.}(2008){Hughes}, {Wilner}, {Qi}, \&
  {Hogerheijde}}]{hug08}
{Hughes}, A.~M., {Wilner}, D.~J., {Qi}, C., \& {Hogerheijde}, M.~R. 2008, \apj,
  678, 1119

\bibitem[{{Igea} \& {Glassgold}(1999)}]{ige99}
{Igea}, J. \& {Glassgold}, A.~E. 1999, \apj, 518, 848

\bibitem[{{Isella} {et~al.}(2009){Isella}, {Carpenter}, \& {Sargent}}]{ise09}
{Isella}, A., {Carpenter}, J.~M., \& {Sargent}, A.~I. 2009, \apj, 701, 260

\bibitem[{{Isella} {et~al.}(2007){Isella}, {Testi}, {Natta}, {Neri}, {Wilner},
  \& {Qi}}]{ise07}
{Isella}, A., {Testi}, L., {Natta}, A., {Neri}, R., {Wilner}, D., \& {Qi}, C.
  2007, \aap, 469, 213

\bibitem[{{Johansen} \& {Klahr}(2005)}]{joh05}
{Johansen}, A. \& {Klahr}, H. 2005, \apj, 634, 1353

\bibitem[{{Johansen} {et~al.}(2007){Johansen}, {Oishi}, {Low}, {Klahr},
  {Henning}, \& {Youdin}}]{joh07}
{Johansen}, A., {Oishi}, J.~S., {Low}, M.-M.~M., {Klahr}, H., {Henning}, T., \&
  {Youdin}, A. 2007, \nat, 448, 1022

\bibitem[{{Kastner} {et~al.}(1999){Kastner}, {Huenemoerder}, {Schulz}, \&
  {Weintraub}}]{kas99}
{Kastner}, J.~H., {Huenemoerder}, D.~P., {Schulz}, N.~S., \& {Weintraub}, D.~A.
  1999, \apj, 525, 837

\bibitem[{{Kastner} {et~al.}(1997){Kastner}, {Zuckerman}, {Weintraub}, \&
  {Forveille}}]{kas97}
{Kastner}, J.~H., {Zuckerman}, B., {Weintraub}, D.~A., \& {Forveille}, T. 1997,
  Science, 277, 67

\bibitem[{{Lay} {et~al.}(1994){Lay}, {Carlstrom}, {Hills}, \&
  {Phillips}}]{lay94}
{Lay}, O.~P., {Carlstrom}, J.~E., {Hills}, R.~E., \& {Phillips}, T.~G. 1994,
  \apjl, 434, L75

\bibitem[{{Lynden-Bell} \& {Pringle}(1974)}]{lyn74}
{Lynden-Bell}, D. \& {Pringle}, J.~E. 1974, \mnras, 168, 603

\bibitem[{{Mamajek}(2005)}]{mam05}
{Mamajek}, E.~E. 2005, \apj, 634, 1385

\bibitem[{{Matsumura} {et~al.}(2009){Matsumura}, {Pudritz}, \&
  {Thommes}}]{mat09}
{Matsumura}, S., {Pudritz}, R.~E., \& {Thommes}, E.~W. 2009, \apj, 691, 1764

\bibitem[{{Mundy} {et~al.}(1993){Mundy}, {McMullin}, {Grossman}, \&
  {Sandell}}]{mun93}
{Mundy}, L.~G., {McMullin}, J.~P., {Grossman}, A.~W., \& {Sandell}, G. 1993,
  Icarus, 106, 11

\bibitem[{{Nelson} \& {Papaloizou}(2003)}]{nel03}
{Nelson}, R.~P. \& {Papaloizou}, J.~C.~B. 2003, \mnras, 339, 993

\bibitem[{{Nelson} \& {Papaloizou}(2004)}]{nel04}
---. 2004, \mnras, 350, 849

\bibitem[{{Pani{\'c}} {et~al.}(2008){Pani{\'c}}, {Hogerheijde}, {Wilner}, \&
  {Qi}}]{pan08}
{Pani{\'c}}, O., {Hogerheijde}, M.~R., {Wilner}, D., \& {Qi}, C. 2008, \aap,
  491, 219

\bibitem[{{Papaloizou} \& {Nelson}(2003)}]{pap03}
{Papaloizou}, J.~C.~B. \& {Nelson}, R.~P. 2003, \mnras, 339, 983

\bibitem[{{Papaloizou} {et~al.}(2004){Papaloizou}, {Nelson}, \&
  {Snellgrove}}]{pap04}
{Papaloizou}, J.~C.~B., {Nelson}, R.~P., \& {Snellgrove}, M.~D. 2004, \mnras,
  350, 829

\bibitem[{{Pessah} {et~al.}(2007){Pessah}, {Chan}, \& {Psaltis}}]{pes07}
{Pessah}, M.~E., {Chan}, C., \& {Psaltis}, D. 2007, \apjl, 668, L51

\bibitem[{{Pi{\'e}tu} {et~al.}(2007){Pi{\'e}tu}, {Dutrey}, \&
  {Guilloteau}}]{pie07}
{Pi{\'e}tu}, V., {Dutrey}, A., \& {Guilloteau}, S. 2007, \aap, 467, 163

\bibitem[{{Qi} {et~al.}(2004){Qi}, {Ho}, {Wilner}, {Takakuwa}, {Hirano},
  {Ohashi}, {Bourke}, {Zhang}, {Blake}, {Hogerheijde}, {Saito}, {Choi}, \&
  {Yang}}]{qi04}
{Qi}, C., {Ho}, P.~T.~P., {Wilner}, D.~J., {Takakuwa}, S., {Hirano}, N.,
  {Ohashi}, N., {Bourke}, T.~L., {Zhang}, Q., {Blake}, G.~A., {Hogerheijde},
  M., {Saito}, M., {Choi}, M., \& {Yang}, J. 2004, \apjl, 616, L11

\bibitem[{{Qi} {et~al.}(2008){Qi}, {Wilner}, {Aikawa}, {Blake}, \&
  {Hogerheijde}}]{qi08}
{Qi}, C., {Wilner}, D.~J., {Aikawa}, Y., {Blake}, G.~A., \& {Hogerheijde},
  M.~R. 2008, \apj, 681, 1396

\bibitem[{{Qi} {et~al.}(2006){Qi}, {Wilner}, {Calvet}, {Bourke}, {Blake},
  {Hogerheijde}, {Ho}, \& {Bergin}}]{qi06}
{Qi}, C., {Wilner}, D.~J., {Calvet}, N., {Bourke}, T.~L., {Blake}, G.~A.,
  {Hogerheijde}, M.~R., {Ho}, P.~T.~P., \& {Bergin}, E. 2006, \apjl, 636, L157

\bibitem[{{Ratzka} {et~al.}(2007){Ratzka}, {Leinert}, {Henning}, {Bouwman},
  {Dullemond}, \& {Jaffe}}]{rat07}
{Ratzka}, T., {Leinert}, C., {Henning}, T., {Bouwman}, J., {Dullemond}, C.~P.,
  \& {Jaffe}, W. 2007, \aap, 471, 173

\bibitem[{{Sano} {et~al.}(2000){Sano}, {Miyama}, {Umebayashi}, \&
  {Nakano}}]{san00}
{Sano}, T., {Miyama}, S.~M., {Umebayashi}, T., \& {Nakano}, T. 2000, \apj, 543,
  486

\bibitem[{{Shakura} \& {Sunyaev}(1973)}]{sha73}
{Shakura}, N.~I. \& {Sunyaev}, R.~A. 1973, \aap, 24, 337

\bibitem[{{Stone} {et~al.}(1996){Stone}, {Hawley}, {Gammie}, \&
  {Balbus}}]{sto96}
{Stone}, J.~M., {Hawley}, J.~F., {Gammie}, C.~F., \& {Balbus}, S.~A. 1996,
  \apj, 463, 656

\bibitem[{{van den Ancker} {et~al.}(1998){van den Ancker}, {de Winter}, \&
  {Tjin A Djie}}]{anc98}
{van den Ancker}, M.~E., {de Winter}, D., \& {Tjin A Djie}, H.~R.~E. 1998,
  \aap, 330, 145

\bibitem[{{Webb} {et~al.}(1999){Webb}, {Zuckerman}, {Platais}, {Patience},
  {White}, {Schwartz}, \& {McCarthy}}]{web99}
{Webb}, R.~A., {Zuckerman}, B., {Platais}, I., {Patience}, J., {White}, R.~J.,
  {Schwartz}, M.~J., \& {McCarthy}, C. 1999, \apjl, 512, L63

\end{thebibliography}

\end{document}